\documentclass{sig-alternate-per}

\usepackage{graphicx}
\usepackage{subfig}
\usepackage{amsmath}
\usepackage{amssymb}
\usepackage{amsfonts,dsfont}
\usepackage{times}
\usepackage{cite}
\usepackage{url}
\usepackage{hyperref}
\usepackage{breakurl}
\usepackage{caption}
\usepackage{verbatim}
\usepackage{algorithm}
\usepackage{algorithmic}
\usepackage{color}

\DeclareCaptionType{copyrightbox}

\newcommand{\AC}{ {\cal C} }

\newcommand{\AI}{ {\cal I} }

\newcommand{\AL}{ {\cal L} }

\newtheorem{theorem}{ {\bf \hspace{-0.18in} Theorem}}
\newtheorem{lemma}{{\bf\hspace{-0.18in} Lemma}}
\newtheorem{corollary}{ {\bf\hspace{-0.18in} Corollary}}
\def\done{\hspace*{\fill} \rule{1.8mm}{2.5mm} }

\newcommand {\bX}{\mbox{\boldmath $X$}}
\newcommand {\bn}{\mbox{\boldmath $n$}}
\newcommand {\bs}{\mbox{\boldmath $s$}}
\newcommand {\bQ}{\mbox{\boldmath $Q$}}

\begin{document}

\title{Stochastic Modeling of Large-Scale Solid-State Storage Systems:
   Analysis, Design Tradeoffs and Optimization}

\numberofauthors{1}

\author{
\alignauthor Yongkun Li, Patrick P.C. Lee, John C.S. Lui \\
    \affaddr{The Chinese University of Hong Kong}\\
    \email{yongkunlee@gmail.com, \{pclee,cslui\}@cse.cuhk.edu.hk}
}

\maketitle

\begin{abstract}
Solid state drives (SSDs) have seen wide deployment in mobiles, desktops,
and data centers due to their high I/O performance and low energy
consumption. As SSDs write data out-of-place, garbage collection (GC) is
required to erase and reclaim space with invalid data.  However, GC poses
additional writes that hinder the I/O performance, while SSD blocks can
only endure a finite number of erasures.  Thus, there is a
performance-durability tradeoff on the design space of GC.  To
characterize the optimal tradeoff, this paper formulates an analytical
model that explores the full optimal design space of any GC algorithm.  We
first present a stochastic Markov chain model that captures the I/O
dynamics of large-scale SSDs, and adapt the mean-field approach to derive
the asymptotic steady-state performance.  We further prove the model
convergence and generalize the model for all types of workload. Inspired
by this model, we propose a {\em randomized greedy algorithm (RGA)} that
can operate along the optimal tradeoff curve with a tunable parameter.
Using trace-driven simulation on DiskSim with SSD add-ons, we demonstrate
how RGA can be parameterized to realize the performance-durability
tradeoff.
\end{abstract}


\section{Introduction}
\label{sec: introduction}

The increasing adoption of solid-state drives (SSDs) is
revolutionizing storage architectures.  Today's SSDs are mainly
built on NAND flash memory, and provide several attractive features:
high performance in  I/O throughput, low energy consumption, and
high reliability due to their shock resistance property.  As the SSD
price per gigabyte decreases \cite{SSDPriceDecrease},
not only desktops are replacing traditional hard-disk drives (HDDs) with SSDs,
but there is a growing trend of using SSDs in data centers
\cite{SSDinDataCenter2, SSDinDataCenter3}.

SSDs have inherently different I/O characteristics from traditional
HDDs. An SSD is organized in {\em blocks}, each of which usually
contains 64 or 128 {\em pages} that are typically of size 4KB each.
It supports three basic operations: {\em read}, {\em write}, and
{\em erase}.  The read and write operations are performed in a unit
of page, while the erase operation is performed in the block level.
After a block is erased, all pages of the block become {\em clean}.
Each write can only operate on a clean page; when a clean page is
written, it becomes a {\em valid} page.  To improve the write
performance, SSDs use the {\em out-of-place write} approach. That
is, to update data in a valid page, the new data is first written to
a different clean page, and the original page containing old data is
marked {\em invalid}. Thus, a block may contain a mix of clean
pages, valid pages, and invalid pages.

The unique I/O characteristics of SSDs pose different design
requirements from those in HDDs.  Since each write must operate on a
clean page, {\em garbage collection (GC)} must be employed to
reclaim invalid pages. GC can be triggered, for example, when the
number of clean pages drops below a predefined threshold. During GC,
some blocks are chosen to be erased, and all valid pages in an
erased block must first be written to a different free block prior
to the erasure.  Such additional writes  introduce performance
overhead to normal read/write operations. To maintain high
performance, one design requirement of SSDs is to minimize the {\em
cleaning cost}, such that a GC algorithm chooses blocks containing
as few valid pages as possible for reclamation.

However, SSDs only allow each block to tolerate  a limited number of
erasures before becoming unusable.  For instance, the number is
typically 100K for single-level cell (SLC) SSDs and 10K for
multi-level cell (MLC) SSDs \cite{Chen09}.  With more bits being
stored in a flash cell and smaller feature size of flash cells, the
maximum number of erasures tolerable by each block further
decreases, for example, to several thousands or even several
hundreds for the latest 3-bits MLC SSDs \cite{FAST12bleakfeature}.
Thus, to maintain high durability, another design requirement of
SSDs is to maximize {\em wear-leveling} in GC, such that all blocks
should have similar numbers of erasures over time so as to avoid any
``hot'' blocks being worn out soon.

Clearly, there is a performance-durability tradeoff in the GC design
space. Specifically, a GC algorithm with a low cleaning cost may not
achieve efficient wear-leveling, or vice versa.  Prior work
(e.g., \cite{Agrawal08}) addressed the tradeoff, but the study is mainly based
on simulations.  From the viewpoints of SSD practitioners, it
remains an open design issue of how to choose the ``best''
parameters of a GC algorithm to adapt to different tradeoff
requirements for different application needs.  However,
understanding the performance-durability tradeoff is non-trivial,
since it depends on the I/O dynamics of an SSD and the dynamics
characterization becomes complicated with the increasing numbers of
blocks/pages of the SSD.   This motivates us to formulate a
framework that can efficiently capture the optimal design space of
GC algorithms and guide the choices of parameterizing a GC algorithm
to fit any tradeoff requirement.



In this paper, we develop an analytical model that characterizes the I/O
dynamics of an SSD and the optimal performance-durability tradeoff of a
GC algorithm.
Using our model as a baseline, we propose a {\em tunable} GC algorithm for
different performance-durability tradeoff requirements.  To summarize, our
paper makes the following contributions:
\begin{itemize}
\item We formulate a stochastic Markov chain model that captures the I/O
    dynamics of an SSD.  Since the state space of our stochastic model
    increases with the SSD size, we adapt the {\em mean field technique}
    \cite{meanfield_boundec, meanfield_loadbalance} to make the model
    tractable.  We formally prove the convergence results under the
    uniform workload to enable us to analyze the steady-state performance
    of a GC algorithm.  We also discuss how our system model can be
    extended for a general workload.
\item
We identify the optimal extremal points that correspond to the minimum
cleaning cost and the maximum wear-leveling, as well as the optimal tradeoff
curve of cleaning cost and wear-leveling that enables us to explore the
{\em full} optimal design space of the GC algorithms.
\item
Based on our analytical model, we propose a novel GC algorithm called the
{\em randomized greedy algorithm (RGA)} that can be tunable to operate along
the optimal tradeoff curve.  RGA also introduces low RAM usage and low
computational cost.
\item To address the practicality of our work, we conduct
extensive simulations using the DiskSim simulator
\cite{disksim} with SSD extensions \cite{Agrawal08}.  We first
validate via synthetic workloads that our model efficiently
characterizes the asymptotic  steady-state performance.
Furthermore, we consider real-world workload traces and use
trace-driven simulations to study the performance tradeoff and
versatility of RGA.
\end{itemize}

The rest of the paper proceeds as follows. In \S\ref{sec: stochastic}, we
propose a Markov model to capture the system dynamics of an SSD and conduct
the mean field analysis.  We formally prove the convergence, and further
extend the model for a general workload. In \S\ref{sec: design space}, we
study the design tradeoff between cleaning cost and wear-leveling of GC
algorithms. In \S\ref{sec: algorithm}, we propose RGA and analyze its
performance. In \S\ref{sec: numerical}, we validate our model via
simulations. In \S\ref{sec: simulation}, we present the trace-driven
simulation results. In \S\ref{sec: related}, we review related work, and
finally in \S\ref{sec: conclusion}, we conclude the paper.


\section{System Model}
\label{sec: stochastic}

We formulate a Markov chain model to characterize the I/O dynamics of an SSD
under the read, write, and GC operations.  We then analyze the model via the
{\em mean field technique} when the SSD scales with the increasing number of
blocks or storage capacity.

\subsection{Markov Chain Model Formulation}
\label{subsec: markov}

Our model considers an SSD with $N$ blocks of $k$ pages each, where the
typical value of $k$ is 64 or 128 for today's commonly used SSDs.  Since SSDs
use the out-of-place write approach (see \S\ref{sec: introduction}), a write
to a logical page may reflect on any physical page. Therefore, SSDs implement
{\em address mapping} to map a logical page to a physical page.  Address
mapping is maintained in the software {\em flash translation layer (FTL)} in
the SSD controller.  It can be implemented in block level \cite{Qin10}, page
level \cite{Gupta09}, or hybrid form \cite{Chung06,Lee07,Park08}.  A survey
of the FTL design including the address mapping mechanisms can be found in
\cite{Chung09}.  In this paper, our model abstracts out the complexity due to
address mapping; specifically, we focus on the physical address space and
directly characterize the I/O dynamics of physical blocks.

Recall from \S\ref{sec: introduction} that a page can be in one of the three
states: {\em clean}, {\em valid} or {\em invalid}.
We classify each block into a different type based on the number of valid
pages containing in the block.  Specifically, a block of type $i$ contains
exactly $i$ valid pages.  Since each block has $k$ pages, a block can be of
one of the $k\!+\!1$ types (i.e., from 0 to $k$ valid pages).  If a block is
of type $i$, then we say it is in state $i$.  Let $X_n(t)$ denote the state of
block $n \in \{1,...,N\}$ at time $t$.  Then the state descriptor for the
whole SSD is
\begin{equation}
\bX^N(t)=(X_1(t), X_2(t), \ldots, X_N(t)),
\label{equation: state 1}
\end{equation}
where $X_i(t)\! \in\! \{0,1,...,k\}$.  Thus, the state space cardinality is
$(k\!+\!1)^N$.  To facilitate our analysis under the large system regime (as
we will show later), we transform the above state descriptor to:
\begin{equation}
\bn^N(t) = ( n_0(t), n_1(t), \ldots, n_k(t) ),
\label{equation: state 2}
\end{equation}
where $n_i(t) \!\in\! \{0,1,...,N\}$ denotes the number of type $i$ blocks in
the SSD.  Clearly, we have $\sum_{j=0}^k \!n_j(t) \!= \!N$, and the state
space cardinality is $\binom{N+k}{k}$.

We first describe how different I/O requests affect the system dynamics of
an SSD from the perspective of physical blocks.  The I/O requests  can be
classified into four types: (1) read a page, (2) perform GC on a block, (3)
program (i.e., write) new data to a page, and (4) invalidate a page. First,
read requests do not change $\bn^N(t)$.  For GC, the SSD selects a block,
writes all valid pages of that block to a clean block, and finally erases
the selected block. Thus, GC requests do not change the state of $\bn^N(t)$
either.   On the other hand, for the program and invalidate requests, if the
corresponding block is of type $i$, it will move from state $i$ to state
$i\!+\!1$ and to state $i\!-\!1$, respectively.

\begin{figure}[htb]
\centering
\includegraphics[width=\linewidth]{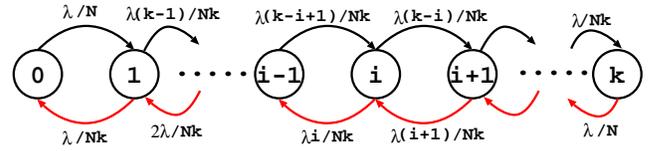}\\
\caption{State transition of a block in an SSD.}
\label{fig: state transition}
\end{figure}

We now describe the state transition of a block in an SSD. Since the read
and GC requests do not change $\bn^N(t)$, we only need to model the program
and invalidate requests.  Suppose that  the program and invalidate requests
arrive as a Poisson process with rate $\lambda$. Also, suppose that the
workload is {\em uniform}, such that all pages in the SSD will have an equal
probability of being accessed (in \S\ref{sec: general}, we extend our model
for a general workload). The assumption of the uniform workload implies that
(1) each block has the same probability $1/N$ of being accessed, (2) the
probability of invalidating one page is proportional to the number of valid
pages of the corresponding block, and (3) the probability of programming a
page is proportional to the total number of invalid and clean pages of the
corresponding block.  Thus, if the requested block is of type $i$, then the
probability of invalidating one page of the block is $\frac{i}{k}$, and that
of programming one page in the block is $\frac{k-i}{k}$. Figure~\ref{fig:
state transition} illustrates the state transitions of a single block in an
SSD under the program and invalidate requests. If a block is in state $i$,
the program and invalidate requests will move it to state $i\!+\!1$ at rate
$\frac{\lambda(k-i)}{Nk}$ and to state $i\!-\!1$ at rate $\frac{\lambda
i}{Nk}$, respectively. Note that Figure~\ref{fig: state transition} only
shows the state transition of a {\em particular} block, but not the  whole
SSD.  Specifically,  the state space cardinality of a particular block is
$k+1$ as shown in Figure~\ref{fig: state transition}, while that of the
whole SSD is $\binom{N+k}{k}$  as described by Equation (\ref{equation:
state 2}).

To characterize the I/O dynamics of an SSD, we define the
{\em occupancy measure} $\boldsymbol{M}^N(t)$ as the vector of fraction of
type $i$ blocks at time $t$.  Formally, we have
$$\boldsymbol{M}^N(t)=(M_0(t), M_1(t),..., M_k(t)),$$ where $M_i(t)$ is
\begin{equation}
  M_i(t)=\frac{1}{N}\sum_{n=1}^N \mathbf{1}_{\{X_n(t)=i\}}
   = \frac{n_i(t)}{N}.
  \label{eq: occupancy}
\end{equation}
In other words, $M_i(t)$ is the {\em fraction} of type $i$ blocks in the SSD.
It is easy to see that the occupancy measure $\boldsymbol{M}^N(t)$ is a
homogeneous Markov chain.

We are interested in modeling {\em large-scale} SSDs to understand the
performance implication of any GC algorithms.  By large-scale, we mean that
the number of blocks $N$ of an SSD is large.  For example, for a 256GB SSD
(which is available in many of today's SSD manufactures), we have
$N\!\approx\!1\!\times\!10^6$ and $k=64$ for a page size of 4KB, implying a
{\em huge} state space of $\boldsymbol{M}^N(t)$. Since $\boldsymbol{M}^N(t)$
does not possess any special structure (i.e., matrix-geometric form),
analyzing it can be computationally expensive.

\subsection{Mean Field Analysis}
\label{subsec: mean field} To make our Markov chain model tractable for a
large-scale SSD, we employ the {\em mean field technique}
\cite{meanfield_boundec, meanfield_loadbalance}. The main idea is that the
stochastic process $\boldsymbol{M}^N(t)$ can be solved by a deterministic
process $\boldsymbol{s}(t) = (s_0(t),s_1(t), ..., s_k(t))$ as $N\rightarrow
\infty$, where $s_i(t)$ denotes the fraction of blocks of type $i$ at time
$t$ in the deterministic process.  We call $\boldsymbol{s}(t)$ the {\em mean
field limit}.  By solving the deterministic process $\boldsymbol{s}(t)$, we
can obtain the occupancy measure of the stochastic process
$\boldsymbol{M}^N(t)$.

We introduce the concept of \textit{intensity} denoted by
$\varepsilon(N)$. Intuitively, the probability that a block performs
a state transition per time slot is in the order of
$\varepsilon(N)$.  Under the uniform workload, each block is
accessed with the same probability $1/N$, so we have
$\varepsilon(N)=1/N$.  Now, we re-scale the process
$\boldsymbol{M}^N(t)$ to $\widetilde{\boldsymbol{M}}^N(t)$.

\begin{equation}
    \widetilde{\boldsymbol{M}}^N(t\varepsilon(N)) = \boldsymbol{M}^N(t)
    \quad \forall t \geq 0.
    \label{eq: rescale}
\end{equation}

For simplicity, we drop the notation $t$ when the context is clear.  We now
show how the deterministic process $\boldsymbol{s}(t)$ is related to the
re-scaled process $\widetilde{\boldsymbol{M}}^N(t)$.  The time evolution of
the deterministic process can be specified by the following set of ordinary
differential equations (ODEs):
\begin{equation}
   \begin{aligned}
     \frac{d s_i}{dt} \!=\!& -\lambda s_i \!+\! \lambda \frac{k\!-\!i\!+\!1}{k} s_{i\!-\!1} \!+\! \lambda \frac{i\!+\!1}{k} s_{i\!+\!1}, \quad 1 \!\leq\! i \!\leq\! k\!-\!1 ,\\
      \frac{d s_0}{dt}\!=\!& -\lambda s_0 \!+\! \lambda\frac{1}{k} s_{1}, \\
      \frac{d s_k}{dt}\!=\!& -\lambda s_k \!+\! \lambda \frac{1}{k} s_{k\!-\!1}.
    \end{aligned}
    \label{eq: ODE}
\end{equation}

The idea of the above ODEs is explained as follows. For an SSD with $N$
blocks, we express the expected change in number of blocks of type $i$ over
a small time period of length $dt$ under the re-scaled process
$\widetilde{\boldsymbol{M}}^N(t)$.  This corresponds to the expected change
over the time period of length $Ndt$ under the original process
$\boldsymbol{M}^N(t)$.  During this period (of length $Ndt$), there are
$\lambda (Ndt)$ program/invalidate requests, each of which changes the state
of some type $i$ block to state $i-1$ or state $i+1$ with probability $1/N$.
Since there are a total of $Ns_i$ blocks of type $i$, the expected change
from state $i$ to other states is $\lambda Ndt s_i$.  Using the similar
arguments, the expected change in number of blocks from state $i\!+\!1$ to
state $i$ is $\lambda Ndt \frac{i+1}{k}s_{i+1}$, and that from state $i-1$
to state $i$ is $\lambda Ndt \frac{k-i+1}{k}s_{i-1}$.  Similarly, we can
also specify the expected change in fraction of blocks of type $0$ and type
$k$, and we obtain the ODEs as stated in Equation~(\ref{eq: ODE}).


\subsection{Derivation of the Fixed Point}
\label{subsec: fixed point}

We now derive the {\em fixed point} of the deterministic process in
Equation~(\ref{eq: ODE}).  Specifically, $\boldsymbol{s}(t)$ is said to be a
fixed point if $\boldsymbol{s}(t) = \boldsymbol{\pi}$ implies
$\boldsymbol{s}(t') = \boldsymbol{\pi}$ for all $t' \geq t$. In other words,
the fixed point $\boldsymbol{\pi}$ describes the distribution of different
types of blocks in the steady state.  The necessary and sufficient condition for
$\boldsymbol{\pi}$ to be a fixed point is that $\frac{d\pi_i}{dt}=0$ for all
$i \in \{0,1,\ldots,k\}$.
\begin{theorem}
Equation~(\ref{eq: ODE}) has a unique fixed point $\boldsymbol{\pi}$ given by:
  \begin{equation}
    \pi_i=\frac{\binom{k}{i}}{2^k},\, 0\leq i \leq k.
    \label{eq: fixed point}
  \end{equation}
  \label{theo: fixed point}
\end{theorem}

\noindent {\bf Proof:} First, it is easy to check that
$\boldsymbol{\pi}$ satisfies $\frac{d\pi_i}{dt}\!=\!0$ for $0\!\leq\! i\! \leq \!k$.
Conversely, based on the condition of $\frac{d\pi_i}{dt}=0$ for all $i$,
we have
\begin{equation*}
    \begin{aligned}
     &- \pi_i \!+\!  \frac{k-i+1}{k} \pi_{i-1}
       \!+\!  \frac{i+1}{k} \pi_{i+1} = 0, \quad 1 \leq i \leq k\!-\!1 ,\\
     &- \pi_0 \!+\! \frac{1}{k} \pi_{1} = 0, \\
     &- \pi_k + \frac{1}{k} \pi_{k-1} = 0.
    \end{aligned}
\end{equation*}
By solving these equations, we get
\begin{equation*}
    \pi_i = \binom{k}{i}\pi_k, \textrm{ for } 0 \leq i \leq k.
\end{equation*}
Since $\sum\limits_{i=0}^k \!\pi_i \!\!=\!\! 1$,  the fixed point is derived
as in Equation~(\ref{eq: fixed point}). \done

\subsection{Summary}

We develop a stochastic Markov chain model to characterize the I/O dynamics
of a large-scale SSD system.  Specifically, we solve the stochastic process
with a deterministic process via the mean field technique and identify the
fixed point in the steady state.  We claim that the derivation is accurate
when $N$ is large, as we can formally provide that (i) the stochastic
process converges to the deterministic process as $N\rightarrow\infty$ and
(ii) the deterministic process specified by Equation~(\ref{eq: ODE})
converges  to the unique fixed point $\boldsymbol{\pi}$ as described in
Equation~(\ref{eq: fixed point}).
 We refer readers to Appendix for
the convergence proofs.

Our model enables us to analyze the tradeoff between cleaning cost
and wear-leveling of GC algorithms. As shown in \S\ref{sec: design
space},  cleaning cost and wear-leveling can be expressed as
functions of $\boldsymbol{\pi}$.


\subsection{Extensions to General Workload}
\label{sec: general}

Our model thus far focuses on the uniform workload, i.e., all physical pages
have the same probability of being accessed.  For completeness, we now
generalize our model to allow for the general workload, in which
blocks/pages are accessed with respect to some general probability
distribution.   We show how we apply the mean field technique to approximate
the I/O dynamics of an SSD, and we also conduct simulations using synthetic
workloads to validate our approximation (see \S\ref{subsec: validation on
fixed-point}).
		As stated in \S\ref{subsec: markov}, we focus on the program
and invalidate requests, both of which can change the state of a
block in the Markov chain model.  We again assume that the
program/invalidate requests arrive as a Poisson process with rate
$\lambda$.  In particular, to model the general workload, we let
$p_{i,j}$ be the transition probability of a type $i$ block being
transited to state $j$ due to one program/invalidate request. We
have
\begin{equation*}
    \begin{aligned}
      &p_{i,j}=0,  \quad \mbox{ if } j \neq i-1 \mbox{ and } j \neq i+1,\\
      &\sum_i\sum_j p_{i,j}\left( \sum_n \mathbf{1}_{\{X_n(t)=i\}} \right)=1,
    \end{aligned}
\end{equation*}
where $\mathbf{1}_{\{X_n(t)=i\}}$ indicates whether block $n$ is in
state $i$, and thus $\sum_n \mathbf{1}_{\{X_n(t)=i\}}$ represents
the number of blocks in state $i$. The second equation comes from
the fact that each program/invalidate request can only change the
state of one particular block.

In practice,  $p_{i,j}$ (where $j=i-1$ or $j=i+1$) can be estimated
via workload traces.  Specifically, for each request being
processed, one can count the number of blocks in state $i$ (i.e.,
$n_i$) and the number of blocks in state $i$ that change to state
$j$ (i.e., $n_{i,j}$). Then $p_{i,j}$ can be estimated as:
\begin{equation}
    \centering
    p_{i, j} \approx \frac{\sum_{\mbox{\small for each request}} \frac{n_{i,j}}{n_i}  }{\mbox{total number of requests}},
  \label{eq: estimation_tran_prob}
\end{equation}
where $\frac{n_{i,j}}{n_i}$ is the probability that a block transits
from state $i$ to  $j$ in a particular request, and $p_{i, j}$ is
the average over all requests.


We can derive the occupancy measure $\boldsymbol{M}^N(t)$ with a
deterministic process $\boldsymbol{s}(t)$ specified by the following
ODEs:
\begin{equation}
    \begin{aligned}
     \frac{d s_i}{dt} \!\!=& \!-\!\lambda (p_{i,i\!-\!1}\!+\!p_{i,i\!+\!1}) s_i \!+\! \lambda p_{i\!-\!1,i}  s_{i\!-\!1} \!+\! \lambda p_{i\!+\!1,i}  s_{i\!+\!1},
            1 \!\!\leq \!\!i\! \!\leq\!\! k\!\!-\!\!1 ,\\
      \frac{d s_0}{dt}\!=& -\lambda p_{0,1} s_0 \!+\! \lambda p_{1,0} s_{1}, \\
      \frac{d s_k}{dt}\!=& -\lambda p_{k,k-1} s_k \!+\! \lambda p_{k-1,k} s_{k\!-\!1}.
    \end{aligned}
    \label{eq: ODE_general}
\end{equation}
%

We can further derive the fixed point of the deterministic process
$\boldsymbol{s}(t)$ as in Theorem \ref{theo: fixed point general}.
 For the convergence proof, please refer to Appendix.
%
%
\begin{theorem}
Equation~(\ref{eq: ODE_general}) has a unique fixed point $\boldsymbol{\pi}$
given by:
  \begin{equation}
    \begin{aligned}
      & \pi_k= \frac{1}{1+\sum_{i=0}^{k-1}\frac{\prod_{j=k}^{i+1}p_{j,j-1}}{\prod_{j=i}^{k-1}p_{j,j+1}}},\\
      & \pi_i=\frac{\prod_{j=k}^{i+1}p_{j,j-1}}{\prod_{j=i}^{k-1}p_{j,j+1}}\pi_k, \quad \quad 0 \leq i\leq k-1.
    \end{aligned}
    \label{eq: fixed_point_general}
  \end{equation}
  \label{theo: fixed point general}
\end{theorem}
\noindent{\bf Proof:} The derivation is similar to that of
Theorem~\ref{theo: fixed point}.
\done

\section{Design Space of GC Algorithms}
\label{sec: design space}

Using our developed stochastic model, we analyze how we can
parameterize a GC algorithm to adapt to different
performance-durability tradeoffs.  In this section, we formally
define two metrics, namely {\em cleaning cost} and {\em
wear-leveling}, for general GC algorithms.  Both metrics are defined
based on the occupancy measure $\boldsymbol{\pi}$ which we derived
in \S\ref{sec: stochastic}. We identify two optimal extremal points
in GC algorithms.  Finally, we identify the optimal tradeoff curve
that explores the full optimal design space of GC algorithms.

\subsection{Metrics}
\label{subsec: metrics}

We now define the new parameters that are used to characterize a family of
GC algorithms.  When a GC algorithm is executed, it selects a block to
reclaim. Let $w_i \geq 0$ (where $0\le i\le k$) denote the weight of
selecting a  particular type $i$ block (i.e., a block with $i$ valid pages),
 such that the higher the weight $w_i$ is, the more likely each
type $i$ block is chosen to be reclaimed. The weights are chosen with the
following constraint:
\begin{equation}
     \sum_{i=0}^k \frac{w_i}{N} \times n_i =
	 \sum_{i=0}^k w_i \pi_i = 1.
\end{equation}
The above constraint has the following physical meaning. The ratio
$w_i/N$ can be viewed as the probability of selecting a particular
type $i$ block for a GC operation.  Since $n_i$ is the total number
of type $i$ blocks in the system, $w_i \pi_i$ can be viewed as the
probability of selecting {\em any} type $i$ block for a GC
operation.   The summation of $w_i \pi_i$ over all $i$ is equal to
1.  Note that $\pi_i$ is the occupancy measure that we derive in
\S\ref{sec: stochastic}.


We now define two metrics that respectively characterize the performance and
durability of a GC algorithm.
The first metric is called the {\em cleaning cost}, denoted by $\mathcal{C}$,
which is defined as the average number of valid pages contained in the block
that is selected for a GC operation.  This implies that the cleaning cost
reflects the average number of valid pages that need to be written to another
clean block during a GC operation.  The cleaning cost reflects the performance
of a GC algorithm, such that a high-performance GC algorithm should have a low
cleaning cost.  Formally, we have
\begin{equation}
    \mathcal{C} = \sum_{i=0}^k i w_i\pi_i.
    \label{eq: cost}
\end{equation}

The second metric is called the \textit{wear-leveling}, denoted by
$\mathcal{W}$, which reflects how {\em balanced} the blocks are being erased
by a GC algorithm.  To improve the durability of an SSD, each block should
have approximately the same number of erasures.
We use the concept of the fairness index \cite{fainess_index} to define the
degree of wear-leveling $\mathcal{W}$, such that the higher $\mathcal{W}$ is,
the more balanced the blocks are erased. Formally, we have
\begin{equation}
    \mathcal{W} = \frac{(\sum_{i=0}^k\frac{w_i}{N}N\pi_i)^2}{N\sum_{i=0}^k(\frac{w_i}{N})^2N\pi_i} = \left(\sum_{i=0}^kw_i^2\pi_i\right)^{-1}.
    \label{eq: wear-leveling}
\end{equation}
 Note that the rationale of Equation (\ref{eq: wear-leveling})
comes from the fact that $\frac{w_i}{N}$ is the probability of selecting a
{\em particular} type $i$ block, and there are $N\pi_i$ type $i$ blocks in
total. For example, if all $w_i$'s are equal to one, which implies that each
block has the same probability $\frac{1}{N}$ of being selected, then the
wear-leveling index $\mathcal{W}$ achieves its maximum value equal to one as
$\sum_{i=0}^k\pi_i=1$.

The set of $w_i$'s, where $0 \leq i \leq k$, will be our selection
parameters to design a GC algorithm.  In the following, we show how we
select $w_i$'s for different GC algorithms subject to different tradeoffs
between cleaning cost and wear-leveling.   Our results are derived for a
general workload subject to the system state distribution
$\boldsymbol{\pi}$.  Specifically, we also derive the closed-form solutions
under the uniform workload as a case study.

\subsection{GC Algorithm to Maximize Wear-leveling}

Suppose that our goal is to find a set of weight $w_i$'s such that a GC
algorithm maximizes wear-leveling $\mathcal{W}$.
We can formulate the following optimization problem:
\begin{eqnarray}
  \max && \mathcal{W} = \left(\sum_{i=0}^k w_i^2\pi_i \right)^{-1}  \\
  s.t. && \sum_{i=0}^k w_i\pi_i = 1, \nonumber\\
  &&w_i \geq 0.\nonumber
\label{problem: WMP}
\end{eqnarray}

The solution of the above optimization problem is to set $w_i\!=\!1$ for all
$i$, and the corresponding wear-leveling  $\mathcal{W}$ is equal to  $1$.
Note that $\mathcal{W}\leq 1$ as $\sum_{i=0}^k w_i^2\pi_i-(\sum_{i=0}^k
w_i\pi_i)^2=\sum_{i=0}^k w_i^2\pi_i-1\geq 0$, so the above solution is the
optimal solution. The corresponding cleaning cost is $\sum_{i=0}^ki\pi_i$.
In other words, each block has the same probability (i.e., $1/N$) of being
selected for GC. Intuitively, this assignment strategy which maximizes
wear-leveling is the \textit{random algorithm}, in which each block is
uniformly chosen independent of its number of valid pages.

Under the uniform workload, we can compute the closed-form solution of the
cleaning cost $\mathcal{C}$ as:
\begin{eqnarray*}
\mathcal{C} = \sum_{i=0}^kiw_i\pi_i=\sum_{i=0}^ki\frac{\binom{k}{i}}{2^k}=\frac{k}{2}.
\end{eqnarray*}
It implies that a random GC algorithm introduces an average of $k/2$
{\em additional page writes} under the uniform workload.

\subsection{GC Algorithm to Minimize Cleaning Cost}

Suppose now that our goal is to find a set of weight $w_i$'s to minimize the
cleaning cost $\mathcal{C}$, or equivalently, minimize the number of writes
of valid pages during GC.  The optimization formulation is:
\begin{eqnarray}\label{problem: CMP}
  \min && \mathcal{C} =  \sum_{i=0}^k i w_i \pi_i  \\
  s.t. && \sum_{i=0}^k w_i \pi_i = 1, \nonumber\\
  && w_i \geq 0.\nonumber
\end{eqnarray}

The solution of the above optimization problem is to set $w_0=1/\pi_0$ and
$w_i=0$ for all $i>0$ (assuming that there exist some blocks of type 0), and
the cleaning cost $\mathcal{C}$ is equal to 0. Since $\mathcal{C} \geq 0$,
and it is equal to 0 when $w_i=0$ for all $i>0$, the solution is optimal.
The corresponding wear-leveling $\mathcal{W}$ is $\pi_0$.  Intuitively, this
assignment strategy corresponds to the \textit{greedy algorithm}, which
always chooses the block that has the minimum number of valid pages for GC.

Under the uniform workload, the closed-form solution of $\mathcal{W}$
corresponding to the minimum cost is given by:
\begin{eqnarray*}
  \mathcal{W} = \frac{1}{w_0^2\pi_0}=\frac{1}{2^k}.
\end{eqnarray*}
The result shows that the greedy algorithm can significantly degrade
wear-leveling.  For today's commonly used SSDs, the typical value of $k$ is 64
or 128.  This implies that the degree of wear-leveling $\mathcal{W}\approx 0$,
and the durability of the SSD suffers.

\subsection{Exploring the Full Optimal Design Space}
\label{subsec: exploring}


We identify two GC algorithms, namely the random and greedy
algorithms, that correspond to two optimal extremal points of all GC
algorithms.  We now characterize the tradeoff between cleaning cost and
wear-leveling, and identify the {\em full} optimal design space of GC
algorithms. Specifically, we formulate an optimization
problem: {\em given a cleaning cost $\mathcal{C^*}$, what is the
maximum wear-leveling that a GC algorithm can achieve?} Formally, we
express the problem (with respect to $w_i$'s) as follows:
\begin{eqnarray}
\max&&    \mathcal{W} = \left(\sum_{i=0}^k w_i^2\pi_i \right)^{-1}    \\
 s.t.&&    \sum_{i=0}^k w_i \pi_i = 1 ,\nonumber \\
&& \sum_{i=0}^ki w_i \pi_i = \mathcal{C^*},   \nonumber\\
&& w_i \geq 0 .\nonumber
\label{problem: F VS C_max}
\end{eqnarray}
Without loss of generality, we assume that $\pi_i\!>\!0$
$(0\!\leq\! i\!\leq\! k)$.
The solution of the optimization problem is stated in the following
theorem.
\begin{theorem}
Given a cleaning cost $\mathcal{C^*}$, the maximum wear-leveling
$\mathcal{W^*}$ is given by:
\begin{equation}
    \mathcal{W^*} = \left\{\begin{aligned}
            &\pi_0, & \AC^* =0, \\
            &\frac{1}{\sum_{i=0}^{\AI}\gamma_i^2\pi_i}, & 0 < \AC^* < \sum_{i=0}^{k}i\pi_i, \\
            &1, & \AC^* = \sum_{i=0}^{k}i\pi_i,\\
            &\frac{1}{\sum_{i=\AL}^{k}\Gamma_i^2\pi_i}, & \sum_{i=0}^{k}i\pi_i < \AC^* < k ,\\
            & \pi_k, &  \AC^* = k,
          \end{aligned}\right.
    \label{eq: maximum wear-leveling}
\end{equation}
for some constants $\gamma_i$, $\AI$, $\Gamma_i$, and $\AL$.
\label{theo: design_space}
\end{theorem}
\noindent{\bf Proof: } The proof is in Appendix. We also derive the
constants $\AI$, $\gamma_i$, $\AL$, and $\Gamma_i$. \done

%

\section{Randomized Greedy Algorithm}
\label{sec: algorithm}

In this section, we present a {\em tunable} GC algorithm called the
\textit{randomized greedy algorithm} (RGA) that can operate at any given
cleaning cost $\mathcal{C}^*$ and return the corresponding optimal
wear-leveling $\mathcal{W}^*$; or equivalently, RGA can operate at any point
along the optimal tradeoff curve of $\mathcal{C}^*$ and $\mathcal{W}^*$.

\subsection{Algorithm Details}

Algorithm~\ref{alg:rga} shows the pseudo-code of RGA, which operates as
follows.  Each time when GC is triggered, RGA {\em randomly} chooses $d$ out
of $N$ blocks $b_1, b_2, \cdots, b_d$ as candidates (Step~2).  Let $v(b_i)$
denote the number of valid pages of block $b_i$.  Then RGA selects the block
$b^*$ that has the smallest number of valid pages, or the minimum $v(.)$, to
reclaim (Step~3).  We then invalidate block $b^*$ and move its valid pages to
another clean block (Steps~4-5).  In essence, we define a {\em selection
window} of window size $d$ that defines a random subset of $d$ out of $N$
blocks to be selected.  The window size $d$ is the tunable parameter that
enables us to choose between the random and greedy policies.  Intuitively, the
random selection of $d$ blocks allows us to maximize wear-leveling, while the
greedy selection within the selection window allows us to minimize the
cleaning cost.  Note that in the special cases where $d=1$ (resp.
$d\rightarrow\infty$), RGA corresponds to the random (resp. greedy) algorithm.
\begin{algorithm}
\caption{Randomized Greedy Algorithm (RGA)}
\label{alg:rga}
\begin{algorithmic}[1]
    \IF{garbage collection is triggered}
        \STATE randomly choose $d$ blocks $b_1$, $b_2$, ..., $b_d$;
        \STATE find block $b^* = \min_{v(b_i)}\{b_i: b_i\in \{b_1, b_2, ..., b_d\}\}$;
        \STATE write all valid pages in $b^*$ to another clean block;
        \STATE erase $b^*$;
    \ENDIF
\end{algorithmic}
\end{algorithm}

\subsection{Performance Analysis of RGA}

We now derive the cleaning cost and wear-leveling of RGA.  We first determine
the values of weights $w_i$'s for all $i$.  Recall from \S\ref{subsec: metrics}
that $w_i \pi_i$ represents the probability of choosing any block of type $i$
for GC.
In RGA, a type $i$ block is chosen for GC if and only if the randomly chosen
$d$ blocks all contain at least $i$ valid pages and at least one of them
contains $i$ valid pages. Thus, the corresponding probability $w_i \pi_i$
is $(\sum_{j=i}^k\pi_i)^d-(\sum_{j=i+1}^k\pi_i)^d$.  Note that this expression
assumes that $d$ blocks are chosen uniformly at random from the $N$ blocks
with replacement, while in RGA, these $d$ blocks are chosen uniformly at
random without replacement.  However, we can still use it as approximation
since $d$ is much smaller than $N$ for a large-scale SSD. Therefore, we have
\begin{equation}
    w_i = \frac{(\sum_{j=i}^k\pi_i)^d-(\sum_{j=i+1}^k\pi_i)^d}{\pi_i}.
    \label{eq: assign pi}
\end{equation}
Based on the definitions of cleaning cost $\mathcal{C}$ in
Equation~(\ref{eq: cost}) and wear-leveling $\mathcal{W}$ in
Equation~(\ref{eq: wear-leveling}), we can derive $\mathcal{C}$ and
$\mathcal{W}$:
\begin{eqnarray}
  \mathcal{C} &=&  \sum_{i=0}^k i \left((\sum_{j=i}^k\pi_i)^d-(\sum_{j=i+1}^k\pi_i)^d \right),\label{eq: C randomized greedy}\\
  \mathcal{W} &=& \frac{1}{ \sum\limits_{i=0}^k  \left(\frac{(\sum_{j=i}^k\pi_i)^d-(\sum_{j=i+1}^k\pi_i)^d}{\pi_i}\right)^2\pi_i}.
  \label{eq: W randomized greedy}
\end{eqnarray}
In \S\ref{sec: numerical}, we will show the relationship between
cleaning cost $\mathcal{C}$ and wear-leveling $\mathcal{W}$ based on RGA.  We
show that RGA almost lies on the optimal tradeoff curve of $\mathcal{C}$ and
$\mathcal{W}$.

\subsection{Deployment of RGA}
\label{subsec: deployment}

We now highlight the practical implications when RGA is deployed. RGA is
implemented in the {\em SSD controller} as a GC algorithm.  From our
evaluation (see details in \S\ref{sec: numerical}), a small value of $d$
(which is significantly less than the number of blocks $N$) suffices to make
RGA lie on the optimal tradeoff curve.  This allows RGA to incur low RAM usage
and low computational overhead.  Specifically, RGA only needs to load the meta
information (e.g., number of valid pages) of $d$ blocks into RAM for
comparison.  With a small value of $d$, RGA consumes an only small amount of
RAM space.  Also, RGA only needs to compare $d$ blocks to select the block
with the minimum number of valid pages for GC.  The computational cost is
$O(d)$ and hence very small as well.  Since a practical SSD controller
typically has limited RAM space and computational power, we expect that RGA
addresses the practical needs and can be readily deployed.

We expect that RGA, like other GC algorithms, is only executed periodically or
when the number of free blocks drops below a predefined threshold.  The window
size $d$ can be tunable at different times during the lifespan of the SSD to
achieve different levels of wear-leveling and cleaning cost along the optimal
tradeoff curve.  In particular, we emphasize that the window size
$d$ can be chosen as a {\em non-integer}. In this case, we can simply linearly
extrapolate $d$ between $\lfloor d\rfloor$ and $\lfloor d+1\rfloor$.
Formally, for a given non-integer value $d$, when GC is triggered, RGA can set
the window size as $\lfloor d\rfloor$ with probability $p$ and set the window
size as $\lfloor d+1\rfloor$ with probability $1-p$, where $p$ is given by:
\begin{equation}
d=p\lfloor d\rfloor + (1-p)\lfloor d+1\rfloor.
\label{eq: non_integerd}
\end{equation}
Thus, we can evaluate the values of $w_i$'s as follows:
\begin{equation*}
  w_i(d)=pw_i(\lfloor d\rfloor) + (1-p)w_i(\lfloor d+1\rfloor),
\end{equation*}
based on Equation~(\ref{eq: assign pi}).  The cleaning cost and
wear-leveling of RGA can be computed accordingly via
Equations~(\ref{eq: cost}) and (\ref{eq: wear-leveling})
substituting $w_i(d)$.  More generally, we can obtain the window
size from some probability distribution with the mean value given by
$d$.   This enables us to operate at {\em any} point of the optimal
tradeoff curve.


\section{Model Validation}
\label{sec: numerical}


We thus far formulate an analytical model that characterizes the I/O
dynamics of an SSD, and further propose RGA that can be tuned to realize
different performance-durability tradeoffs.  In this section, we validate
our theoretical results developed in prior sections.  First, we validate via
simulation that our system state derivations in Theorem~\ref{theo: fixed
point general} provide accurate approximation even for a general workload.
Also, we validate that RGA operates along the optimal tradeoff curve
characterized in Theorem~\ref{theo: design_space}.


\subsection{Validation on Fixed-Point Derivations}
\label{subsec: validation on fixed-point}

Recall from \S\ref{sec: stochastic} that we derive, via the mean
field analysis, the fixed-point  $\boldsymbol{\pi}$ for the system
state of our model under both uniform and general workloads.  We now
validate the accuracy of such derivation.  We use the DiskSim
simulator \cite{disksim} with SSD extensions \cite{Agrawal08}.  We
generate synthetic workloads for different read/write patterns to
drive our simulations, and compare the system state obtained by each
simulation with that of our model.

We feed the simulations with three different types of synthetic
workloads: (1) {\sf Random}, (2) {\sf Sequential}, and (3) {\sf
Hybrid}.  Specifically, {\sf Random} means that the starting address
of each I/O request is uniformly distributed in the  logical address
space.  Note that its definition is (slightly) different from that
of the uniform workload used in our model, as the latter directly
considers the requests in the physical address space.  The
logical-to-physical address mapping will be determined by the
simulator.  {\sf Sequential} means that each request starts at the
address which immediately follows the last address accessed by the
previous request. {\sf Hybrid} assumes that there are 50\% of {\sf
Random} requests and 50\% of {\sf Sequential} requests. Furthermore,
for each synthetic workload, we consider both Poisson and non-Poisson
arrivals.  For the former, we assume that the inter-arrival time of requests
follows an exponential distribution with mean 100ms; for the latter, we assume
that the inter-arrival time of requests follow a normal distribution (denoted
by $N(\mu,\sigma^2)$) with mean $\mu$ = 100ms and standard deviation $\sigma$
= 10ms.

Using simulations, we generate 10M requests for each workload and feed them
to a small-scale SSD that contains 8 flash packages with 160 blocks each.
We consider a small-scale SSD (i.e., with a small number of blocks) to make
the SSD converge to an equilibrium state quickly with a sufficient number of
requests; in \S\ref{sec: simulation}, we consider a larger-size SSD.  After
running all 10M requests, we obtain the system state of the SSD for each
workload from our simulation results.  On the other hand, using our model,
we first execute the workload and record the transition probabilities
$p_{i,j}$'s based on Equation~(\ref{eq: estimation_tran_prob}).  We then
compute the system state $\boldsymbol{\pi}$ using Theorem~\ref{theo: fixed
point general} for a general workload (which covers the uniform workload as
well).  We then compare the system states obtained from both the simulations
and model derivations.
\begin{figure*}[!ht]
  \centering
  \subfloat[{\sf Random} + Poisson]{
  \includegraphics[width=0.33\textwidth]{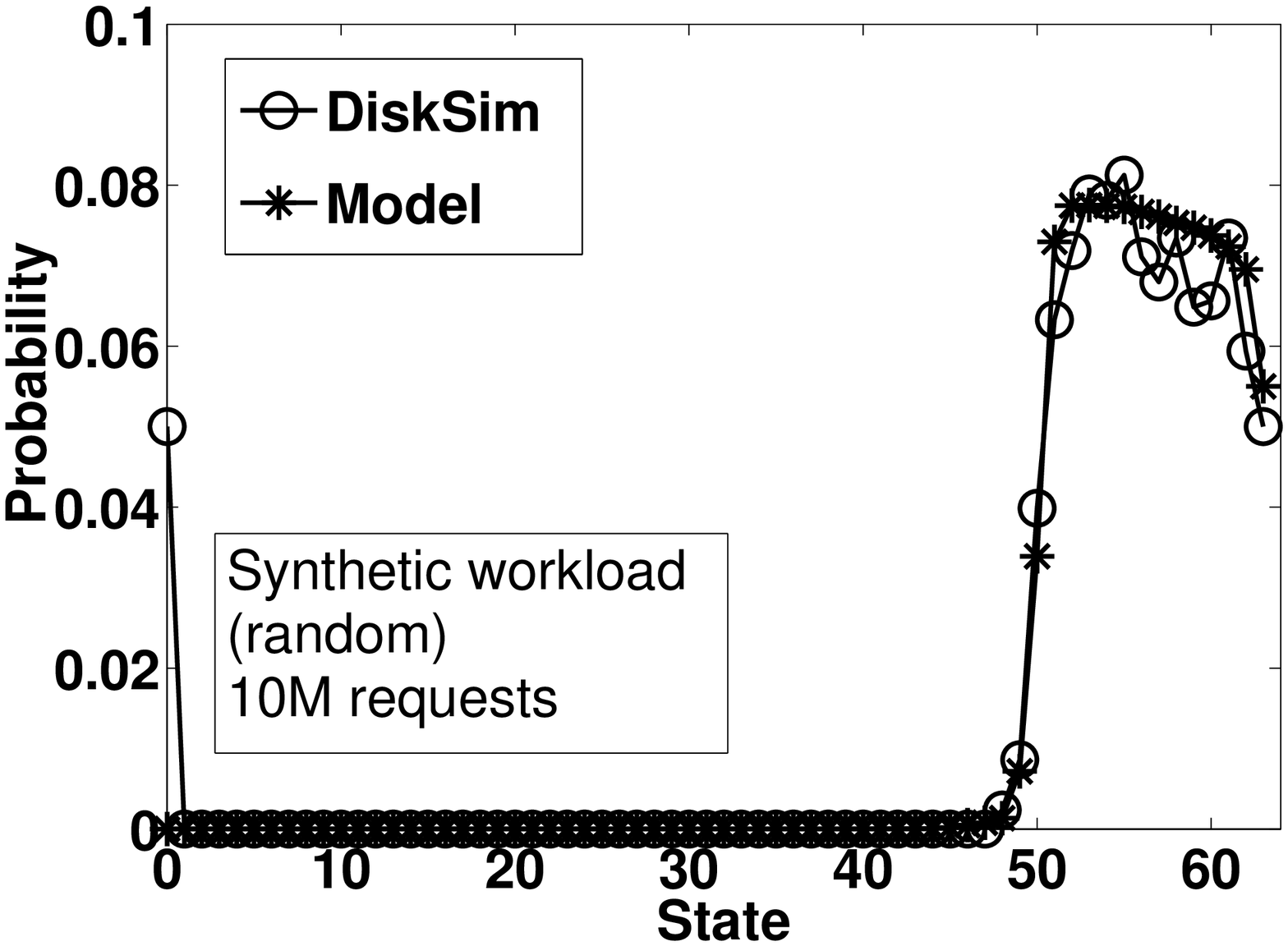}\label{fig: random_poisson}}
  \subfloat[{\sf Sequential} + Poisson]{
  \includegraphics[width=0.33\textwidth]{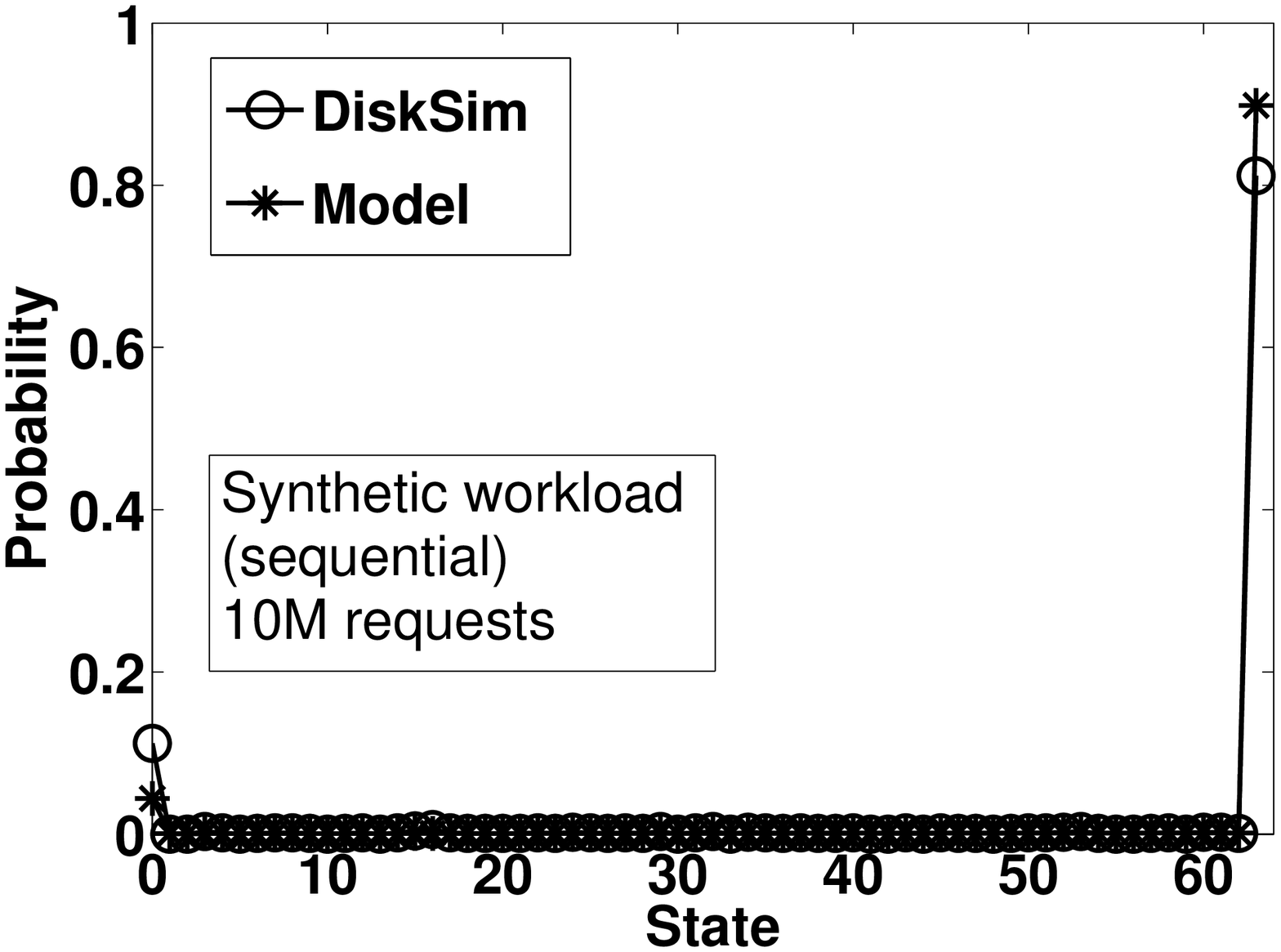}\label{fig: sequential_poisson}}
 \subfloat[{\sf Hybrid} + Poisson]{
  \includegraphics[width=0.33\textwidth]{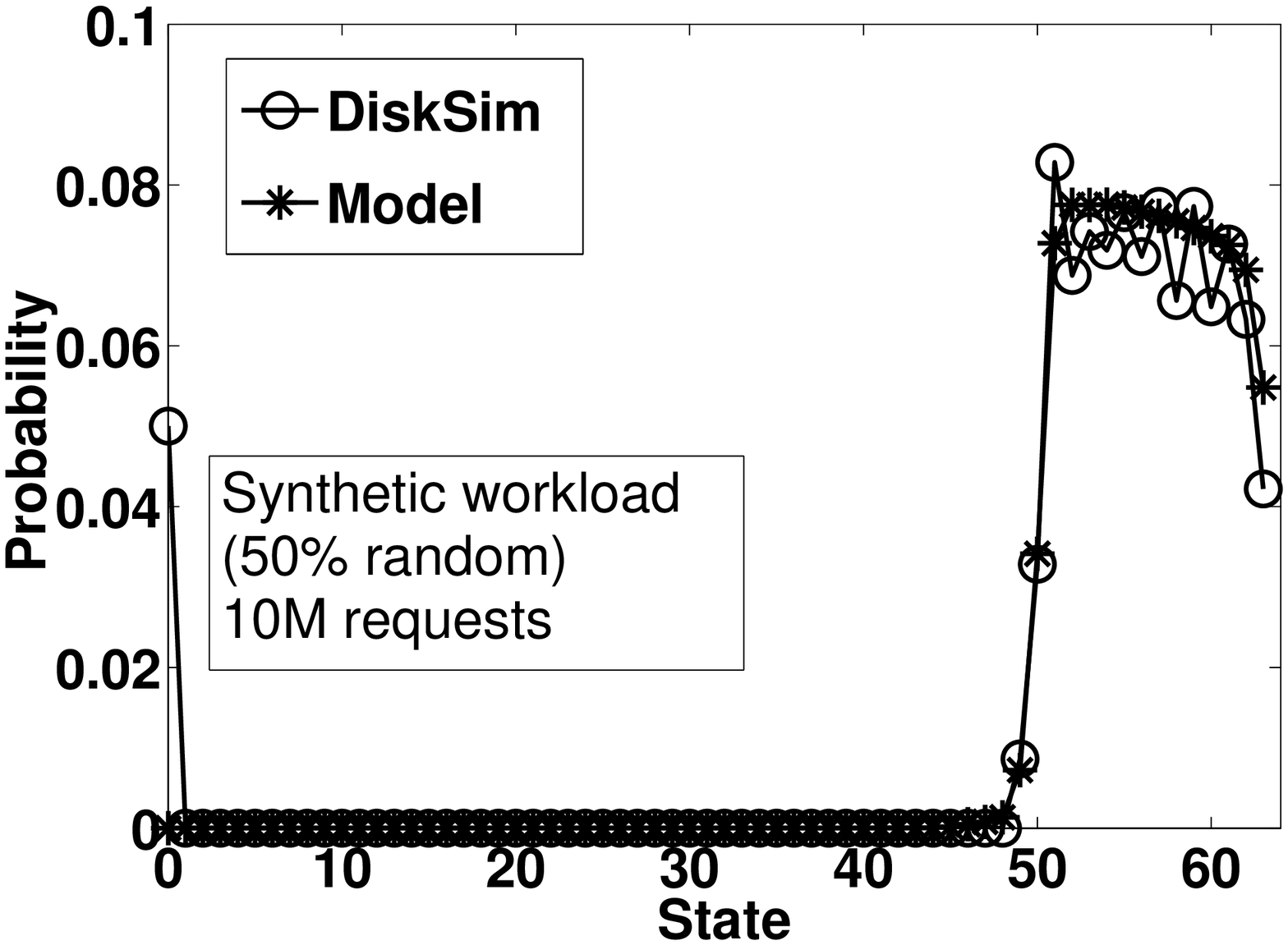}\label{fig: hybrid_poisson}}\\
  \subfloat[{\sf Random} + non-Poisson]{
  \includegraphics[width=0.33\textwidth]{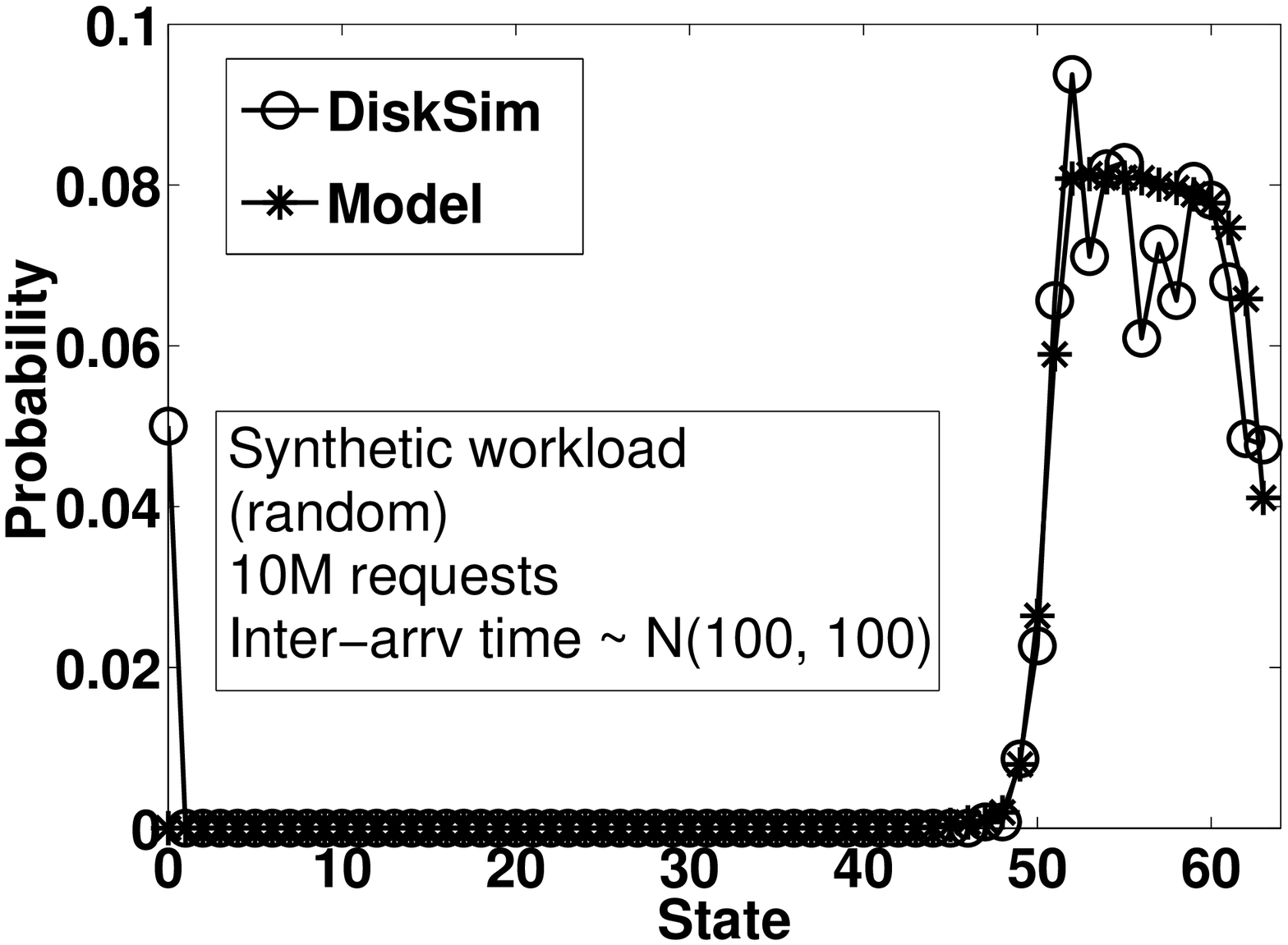}\label{fig: random_nonpoisson}}
  \subfloat[{\sf Sequential} + non-Poisson]{
  \includegraphics[width=0.33\textwidth]{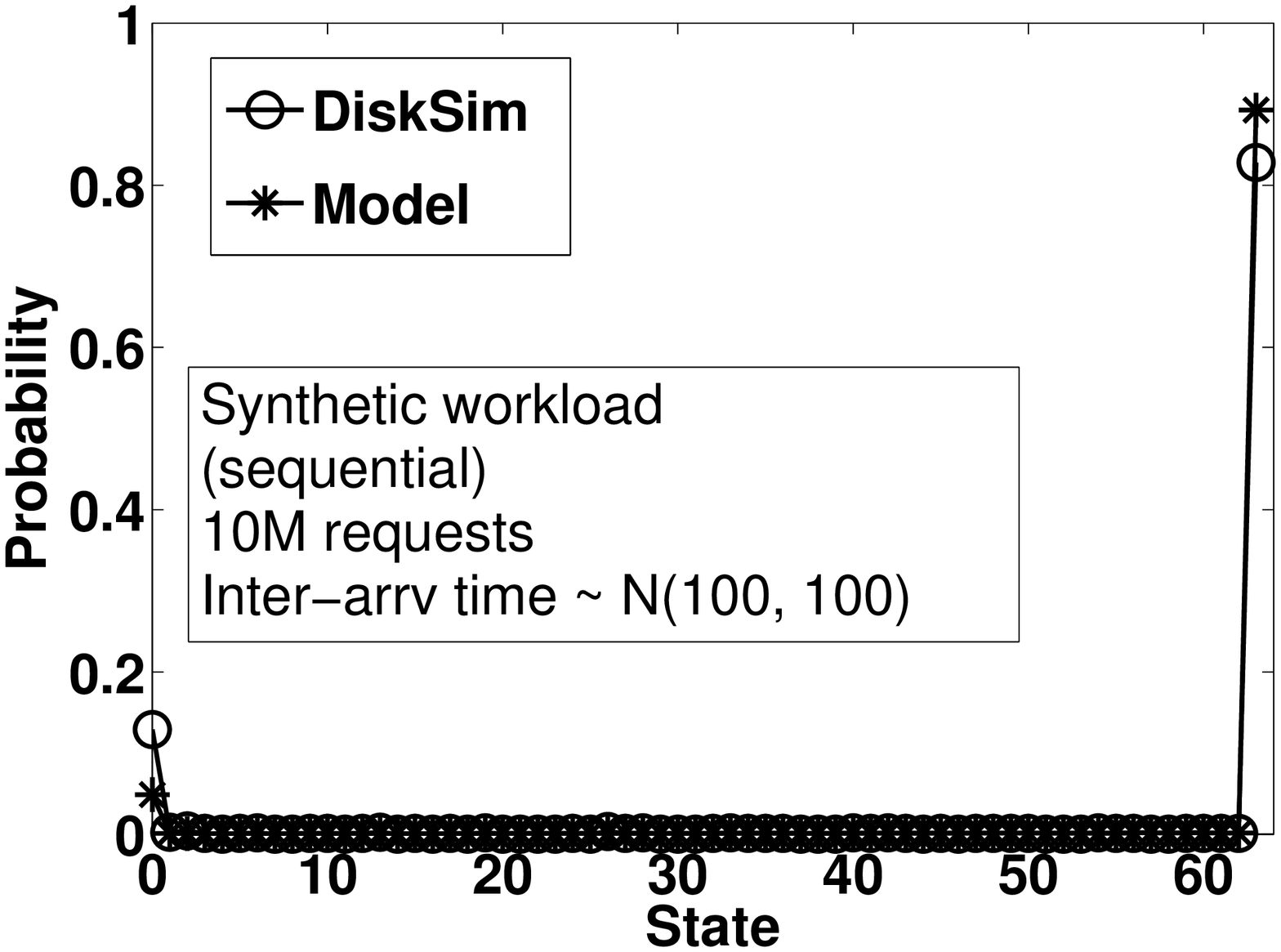}\label{fig: sequential_nonpoisson}}
  \subfloat[{\sf Hybrid} + non-Poisson]{
  \includegraphics[width=0.33\textwidth]{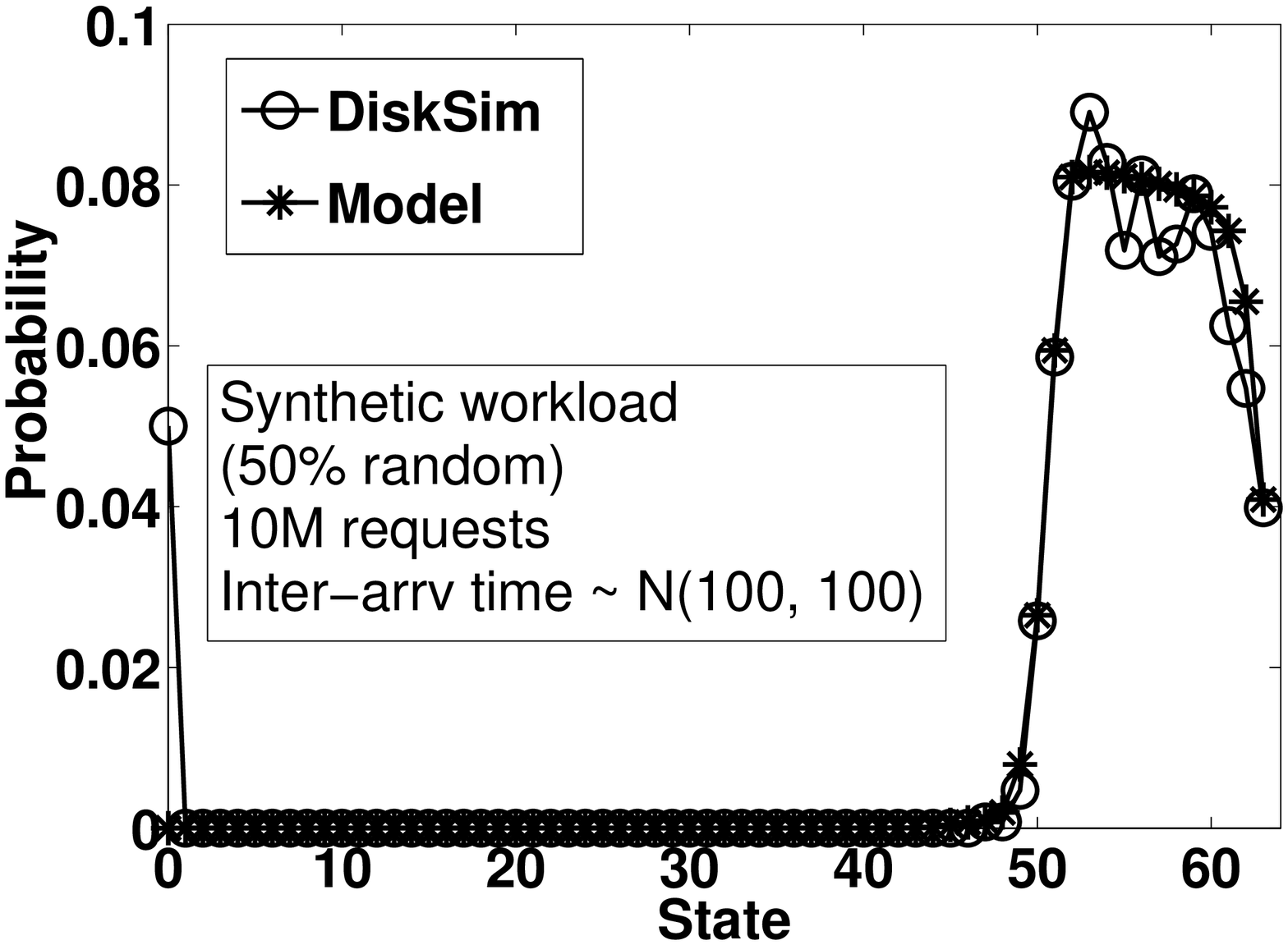}\label{fig: hybrid_nonpoisson}}\\
\caption{Model validation on the system state $\boldsymbol{\pi}$.  In each
sub-figure, the x-axis represents the states (i.e., the number of valid pages  in a block),
and the y-axis indicates the state probabilities.}
\label{fig: validation}
\end{figure*}

Figure~\ref{fig: validation} show the simulation and model results
for the {\sf Random}, {\sf Sequential}, and {\sf Hybrid} workloads,
each associated with either the Poisson or non-Poisson arrivals of requests.
The results show that under different synthetic workloads, our model derived
from the mean field technique can still provide good approximations of the
system state compared with that obtained from the simulations.  Note that we
also observe good approximations even for non-Poisson arrivals of requests.
The results show the robustness of our model in evaluating the system state.

\subsection{Validation on Operational Points of RGA}

In \S\ref{sec: design space}, we characterize the optimal tradeoff curve
between cleaning cost and wear-leveling; in \S\ref{sec: algorithm}, we
present a GC algorithm called RGA that can be tuned by a parameter $d$ to
adjust the tradeoff between cleaning cost and wear-leveling.  We now validate
that RGA can indeed be tuned to operate on the optimal tradeoff curve.

\begin{figure*}[!ht]
  \centering
  \subfloat[Distributions of $\boldsymbol{\pi}$]{
  \includegraphics[width=0.32\textwidth]{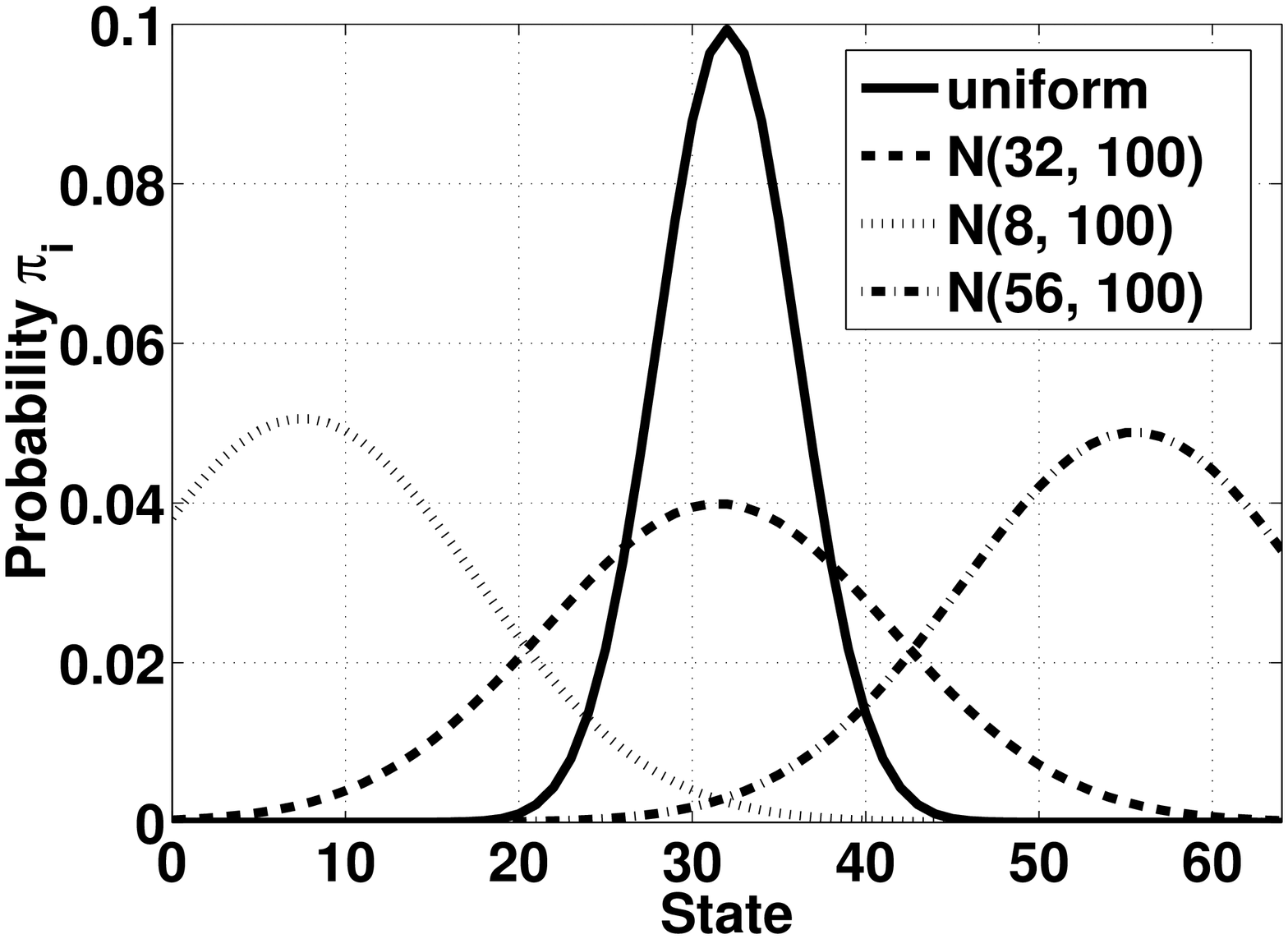}\label{fig: distributionspi}}
  \subfloat[Design space and performance of RGA\newline for different
  integers of $d$ from 1 to 100]{
  \includegraphics[width=0.32\textwidth]{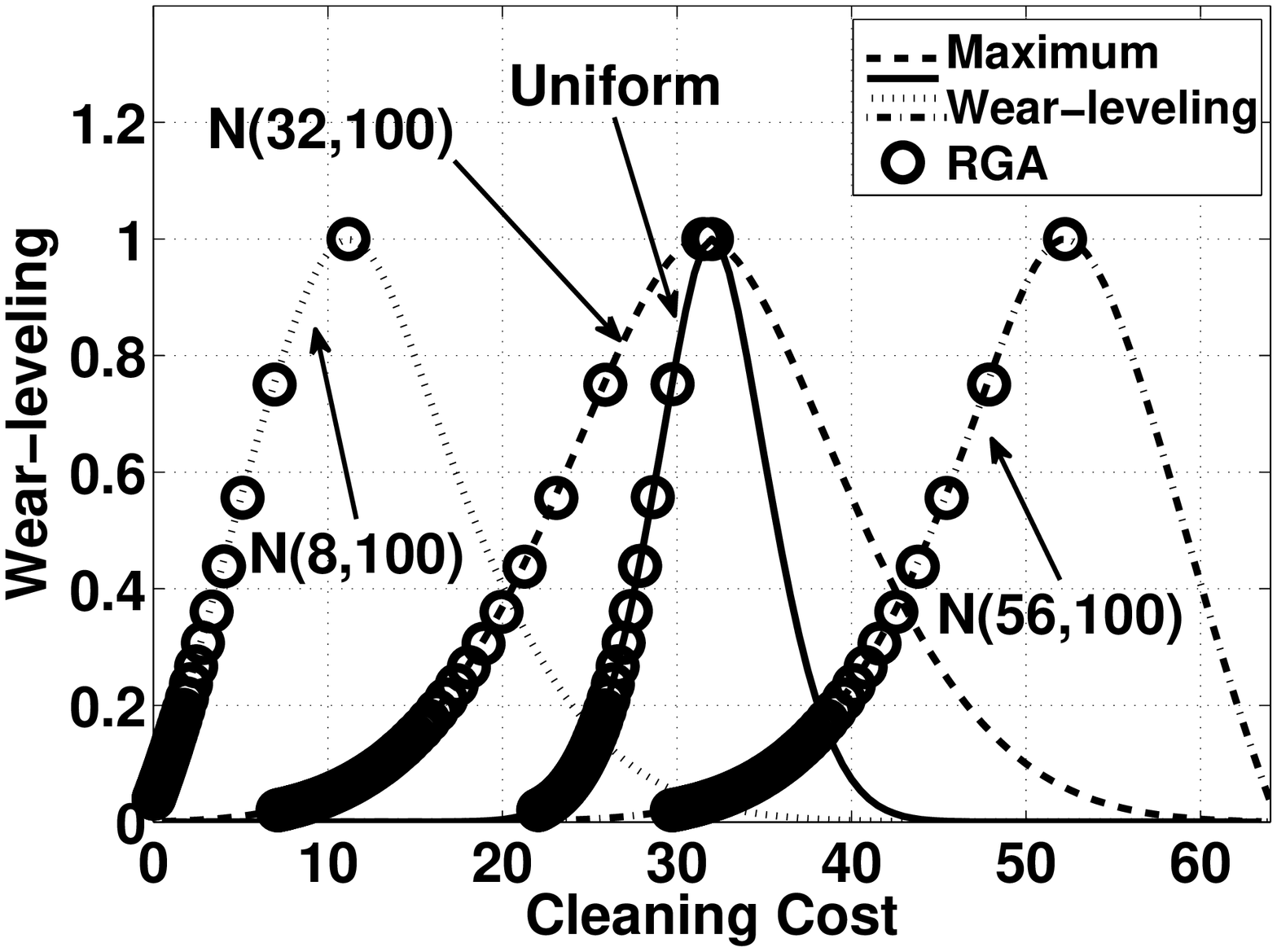}\label{fig: design_space_rga}}
  \subfloat[Design space and performance of RGA for different
  non-integers of $d$ from 1 to 2]{
  \includegraphics[width=0.32\textwidth]{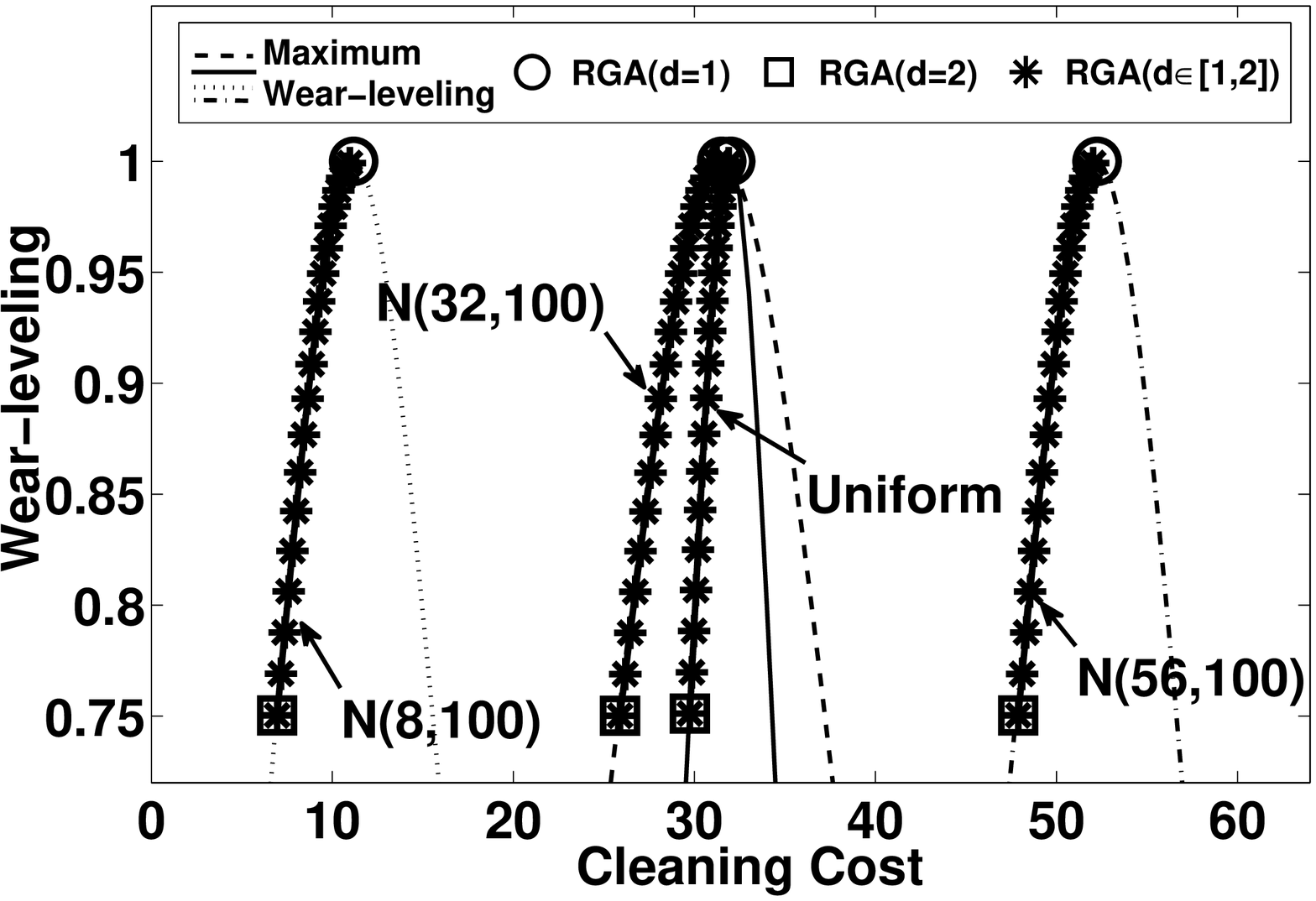}\label{fig: design_space_rga_noni}}
\caption{Full design space and the performance of RGA.}
\label{fig: model_validation_rga}
\end{figure*}

We consider different system state distributions $\boldsymbol{\pi}$
to study the performance of RGA.  We first consider
$\boldsymbol{\pi}$ derived for the uniform workload (i.e.,
Equation~(\ref{eq: fixed point})). We also consider three different
distributions of $\boldsymbol{\pi}$ that are drawn from truncated normal
distributions, denoted by $N(\mu,\sigma^2)$ with mean $\mu$ and standard
deviation $\sigma$. Figure~\ref{fig: model_validation_rga}(a) illustrates the
four system state distributions, where the mean and variance of each truncated
normal distribution are shown in the figure.

For each system state distribution, we compute the maximum wear-leveling
$\mathcal{W^*}$ for each cleaning cost $\mathcal{C^*}$ based on
Theorem~\ref{theo: design_space}. Also, we evaluate the performance of RGA by
varying the window size $d$ from 1 to 100, and obtain the corresponding
cleaning cost and wear-leveling based on Equations~(\ref{eq: C randomized
greedy}) and (\ref{eq: W randomized greedy}).  Here, we only focus on the
integer values of $d$.

Figure~\ref{fig: model_validation_rga}(b) shows the results, in which the four
curves represent the optimal tradeoff curves corresponding to the four
different distributions of $\boldsymbol{\pi}$, while the circles correspond to
the operational points of RGA with different integer values of window size $d$
from 1 to 100.  Note that the maximum wear-leveling corresponds to RGA with
window size $d = 1$ (i.e., the random algorithm).  As the window size
increases, the wear-leveling decreases, while the cleaning cost also
decreases. We observe that RGA indeed operates along the optimal tradeoff
curves with regard to different system state distributions.

It is important to note that we can realize non-integer window sizes to
further fine-tune RGA along the optimal tradeoff curve (see \S\ref{subsec:
deployment}).  To validate, we consider different values of $d$ from 1 to 2,
with step size 0.05, and calculate $d$ via linear extrapolation between 1 and
2.

Figure~\ref{fig: model_validation_rga}(c) shows the results for non-integer
$d$ using different system state distributions.  Here, we zoom into the
wear-leveling values from 0.75 to 1.  Each star corresponds to the RGA with a
non-integer window size obtained by Equation (\ref{eq: non_integerd}). We
observe that RGA can be further fine-tuned to operate along the optimal
tradeoff curves even when $d$ is a non-integer.

\section{Trace-Driven Evaluation}
\label{sec: simulation}

In this section, we evaluate the performance of RGA under more realistic
settings.  Since today's SSD controllers are mainly proprietary firmware, it
is non-trivial to implement GC algorithms inside a real-world SSD controller.
Thus, similar to \S\ref{sec: numerical}, we conduct our evaluation using the
DiskSim simulator \cite{disksim} with SSD extensions \cite{Agrawal08}.  This
time we focus on a large-scale SSD.  We consider several real-world traces,
and evaluate different metrics, including cleaning cost, I/O throughput,
wear-leveling, and durability, for different GC algorithms.  Note
that the cleaning cost and wear-leveling are the metrics considered in the
model, while the I/O throughput and durability are the metrics related to user
experience.

Using trace-driven evaluation, our goal is to demonstrate the effectiveness of
RGA in practical deployment.  We compare different variants of RGA with regard
to different values of window size $d$, as well as the random and greedy
algorithms.  We emphasize that we are {\em not} advocating a particular value
of $d$ for RGA in real deployment; instead, we show how different values of
$d$ can be tuned along the performance-durability tradeoff.

\subsection{Datasets}

We first describe the datasets that drive our evaluation.  Since the read
requests do not influence our analysis, we focus on four real-world traces
that are all write-intensive:
\begin{table*}[!t]
\begin{center}
  \begin{tabular}{||c|c|c|c|c|c||}
  \hline
  \mbox{{\bf Trace}} & {\bf Total \# of requests} & {\bf  Write ratio} & {\bf  Sequential ratio}
     & {\bf  Avg. request size} & {\bf  Avg. inter-arrival time}\\
  \hline \hline
	{\sf Financial} & 4.4 M & 0.7782 & 0.0167& 5.4819 KB& 9.9886 ms\\
	{\sf Webmail} & 7.8 M& 0.8186 & 0.7868 & 4 KB & 222.118 ms\\
	{\sf Online} & 5.7 M& 0.7388& 0.7373& 4 KB & 303.763 ms\\
	{\sf Webmail+Online} & 13.5 M& 0.7849& 0.7597& 4 KB&128.302 ms \\
  \hline
  \end{tabular}
\end{center}
\vspace{-1em}
\caption{Workload statistics of traces.}
\label{table: traces}
\end{table*}
\begin{itemize}
\item
{\sf Financial} \cite{umasstrace}: It is an I/O trace collected from an online
transaction process application running at a large financial institution.
There are two financial traces in \cite{umasstrace},
namely {\sf Financial1.spc} and {\sf Financial2.spc}. Since
{\sf Financial2.spc} is read-dominant, we only use {\sf Financial1.spc} in
this paper.
\item
{\sf Webmail} \cite{fiutrace}: It is an I/O trace that describes the webmail
workload of a university department mail server.
\item
{\sf Online} \cite{fiutrace}: It is an I/O trace that describes the coursework
management workload on Moodle at a university.
\item
{\sf Webmail+Online} \cite{fiutrace}: It is the combination of the I/O traces
of {\sf Webmail} and {\sf Online}.
\end{itemize}
Table~\ref{table: traces} summarizes the statistics of the traces.
The original {\sf Financial} trace in \cite{umasstrace} contains 24
application-specific units (ASUs) of a storage server (denoted by
ASU0 to ASU23). We study the traces of all ASUs except ASU1, ASU3,
and ASU5, whose maximum logical sector numbers go beyond the logical
address space in our configured SSD (see \S\ref{subsec:
configuration}).  The remaining {\sf Financial} trace contains
around 4.4~million I/O requests, in which 77.82\% are write requests
and the remaining are read requests.  Also, 1.67\% of I/O requests
are {\em sequential requests}, each of which has its starting
address immediately following the last address of its prior request.
The average size of each request is 5.4819KB, meaning that most
requests only access one page as the size of one page is configured
as 4KB in the simulation. The average inter-arrival time of two
continuous requests is just around 10 ms.  On the other hand, for
the {\sf Webmail}, {\sf Online} and {\sf Webmail+Online} traces
obtained from \cite{fiutrace}, the write requests account for around
80\% of I/O requests, and over 70\% of I/O requests are sequential
requests. Moreover, all requests in those traces have size 4KB
(i.e., only one page is accessed in each request), and the average
inter-arrival time is much longer than that of the {\sf Financial}
trace.  In summary, the {\sf Financial} trace has the {\em
random-write-dominant} access pattern, while the {\sf Webmail}, {\sf
Online}, and {\sf Webmail+Online} traces have the {\em
sequential-write-dominant} access pattern.

We set the page size of an SSD as 4KB (the default value in most
today's SSDs).  Since the block size considered by these traces is
512~bytes, we align the I/O requests of these traces to be multiples
of the 4KB page size. To enable us to evaluate different GC
algorithms, we need to make the blocks in an SSD undergo a
sufficient number of program-erase cycles. However, these traces may
not be long enough to trigger enough block erasures. Thus, we
propose to {\em replay} a trace; that is, in each replay cycle, we make a copy
of the original trace without changing its I/O patterns, while we only change
the arrival times of the requests by adding a constant value.  In our
simulations, we replay the traces multiple times so that each trace file
contains around 50M I/O requests. Since we replay a trace, we issue the same
write request to a page multiple times, and this keeps invalidating pages due
to out-of-place writes.  Thus, many GC operations will be triggered, and this
enables us to stress-test the cleaning cost and wear-leveling metrics.  We
point out that this replay approach has also been used in the prior SSD work
\cite{MSST11replaytrace}.


\subsection{System Configuration}
\label{subsec: configuration}

Table~\ref{table: parameters} summarizes the parameters that we use
to configure an SSD in our evaluation.  We use the default
configurations from the simulator whose parameters are based on a
common SLC SSD \cite{Chen09}. Specifically, the SSD contains 8 flash
packages, each of which has its own control bus and data bus, so
they can process I/O requests in parallel.  Each flash package
contains 8 planes containing 2048 blocks each. Each block contains
64 pages of size 4KB each.  Therefore, each flash package contains
16384 physical blocks in total and the physical capacity of the SSD
is 32GB. For the timing parameters, the time to read one page from
the flash media to the register in the plane is 25$\mu$s, and the
time of programming one page from the register in the plane to the
flash media is 0.2ms.  For an erase operation, it takes 1.5ms to
erase one block.  The time of transferring one byte through the data
bus line is 0.025$\mu$s.  Since an SSD is usually over-provisioned,
we set the over-provisioning factor as 15\%, which means that the
advertised capacity of an SSD is only 85\% of the physical capacity.
Moreover, we set the threshold of triggering GC as 5\%, meaning that
GC will be triggered when the number of free blocks in the system is
smaller than 5\%. Since flash packages are independent in processing
I/O requests, GC is also triggered independently in each flash
package. In the following, we only focus on a single flash
package and compare the performance of different GC algorithms.

We consider two different initial states of an SSD before we start
our simulations.  The first one is the {\em empty} state, meaning
that the SSD is entirely clean and no data has been stored. The
second one is the {\em full} state, meaning the SSD is fully
occupied with valid data and each logical address is always mapped
to a physical page containing valid data.  Thus, each write request
to a (valid) page will trigger an update operation, which writes the
new data to a clean page and  invalidates the original page. Note
that the full initial state is the default setting in the simulator.
In most of our simulations
(\S\ref{subsec: cleaning cost}-\S\ref{subsec: wear-leveling}), we use the full
initial state as it can be viewed as ``stress-testing'' the I/O performance of
an SSD.  When we study the durability of SSDs (\S\ref{subsec: durability}), we
use the empty initial state as it can be viewed as the state of a brand-new
SSD.

\begin{table}[!t]
\begin{center}
  \begin{tabular}{||c|c||}
  \hline
  \mbox{{\bf Parameter}} & {\bf  Value}  \\
  \hline \hline
  page size & 4KB\\
  \# of pages per block & 64\\
  \# of blocks per package & 16384\\
  \# of packages per SSD & 8\\
  SSD capacity & 32 GB\\
  \hline
  read one page & 0.025ms\\
  write one page & 0.2ms\\
  erase one block & 1.5ms\\
  transfer one byte & 0.000025ms\\
  \hline
  over-provisioning & 15\% \\
  threshold of triggering GC & 5\% \\
  \hline
  \end{tabular}
\end{center}
\vspace{-1em}
  \caption{Configuration parameters.}
\label{table: parameters}
\end{table}

\subsection{Cleaning Cost}
\label{subsec: cleaning cost}

We first evaluate the cleaning cost of different GC algorithms.  In
particular,  we execute the traces with each of the GC algorithms and record
the total number of GC operations and the total number of valid pages which
are written back due to GC. We then derive the cleaning cost as the average
number of valid pages that are written back in each GC operation.

\begin{figure}[!htb]
  \centering
  \includegraphics[width=\linewidth]{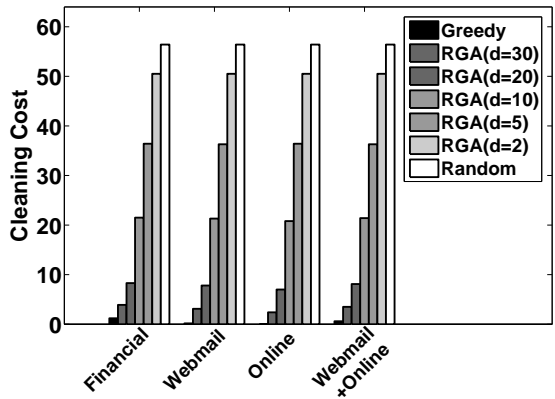}
  \caption{Cleaning cost of different GC algorithms. }
  \label{fig: cleaning_cost}
\end{figure}

Figure \ref{fig: cleaning_cost} shows the simulation results. In
this figure, there are four groups of bars which correspond to the
{\sf Financial},  {\sf Webmail}, {\sf Online}, and {\sf
Webmail+Online} traces, respectively. In each group, there are seven
bars which correspond to the greedy algorithm, random algorithm and
RGA with different window sizes $d$. The vertical axis represents the cleaning
cost that each GC algorithm incurs.  In this simulation,  the simulator starts
from the full initial state. We can see that the greedy algorithm incurs the
smallest cleaning cost that is almost 0, while the random algorithm has the
highest cleaning cost that is close to the total number of pages in each block
(i.e., $k$=64). The intuition is that if the greedy algorithm is used, then
for every GC operation, the block containing the smallest number of valid
pages is reclaimed, which means that it only needs to read out and write back
the smallest number of pages. Therefore, the cleaning cost of the greedy
algorithm should be the smallest among all algorithms. Moreover, RGA provides
a variable cleaning cost between the greedy and random algorithms.

\subsection{Impact on I/O Throughput}
\label{subsec: impact on io}

We now consider the impact of different GC algorithms on the I/O throughput,
using the metric Input/Output Operations Per Second (IOPS).  Note that IOPS is
an {\em indirect} indicator of the cleaning cost.  Specifically, the higher
the cleaning cost, the more pages needed to be moved in each GC operation.
This prolongs the duration of a GC operation, and leads to smaller IOPS as an
I/O request must be queued for a longer time until a GC operation is finished.

\begin{figure}[!htb]
  \centering
  \includegraphics[width=\linewidth]{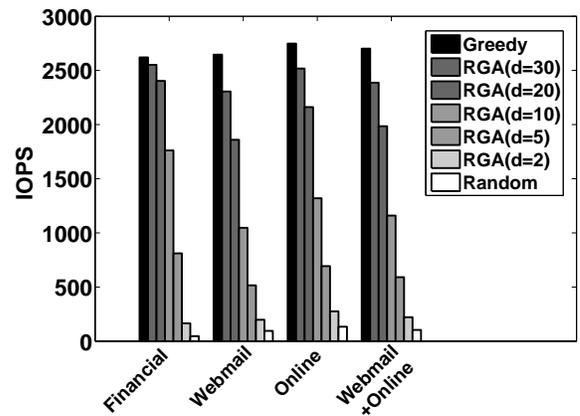}
  \caption{IOPS of different GC algorithms.}
  \label{fig: iops}
\end{figure}

Figure~\ref{fig: iops} shows the IOPS results of different GC algorithms (note
that the simulator starts from the full initial state).  We can see that the
greedy algorithm achieves the highest IOPS, and the random algorithm has the
lowest IOPS, which is less than 5\% of the IOPS achieved by the greedy
algorithm.   The results conform to those in Figure~\ref{fig: cleaning_cost}.
This means that the cleaning cost, the metric that we use in our analytical
model, correctly reflects the resulting I/O performance.  Again, RGA can
provide different I/O throughput results with different values of $d$.

\subsection{Wear-Leveling}
\label{subsec: wear-leveling}

We now evaluate the wear-leveling of different GC algorithms.  In the
simulation, we execute the traces with each of the GC algorithms and record
the number of times that each block has been erased. We then estimate the
probability that each block is chosen for GC and derive the wear-leveling
based on its definition in Equation~(\ref{eq: wear-leveling}).

\begin{figure}[!htb]
  \centering
  \includegraphics[width=\linewidth]{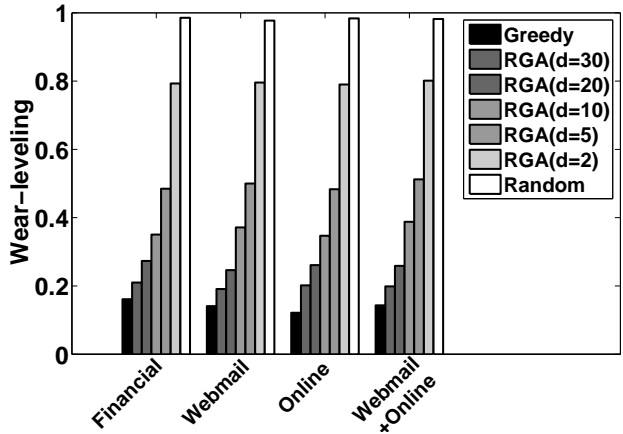}
  \caption{Wear-leveling of different GC algorithms.}
  \label{fig: wear_leveling}
\end{figure}

Figure~\ref{fig: wear_leveling} shows the wear-leveling results.  It
is clear that the random algorithm always achieves the maximum wear-leveling,
which is almost one. This implies that the random algorithm can effectively
balance the numbers of erasures across all blocks.  On the other hand, the
greedy algorithm achieves the minimum wear-leveling which is less than 0.2 for
all traces.
Here, we note that in all traces, our RGA realizes different levels of
wear-leveling between the random and greedy algorithms with different values
of $d$.  In particular, when $d\le 2$, the wear-leveling of RGA is within 80\%
of the maximum wear-leveling of the random algorithm.

\subsection{Impact on Durability}
\label{subsec: durability}

The previous wear-leveling experiment provides insights into the durability
(or lifetime) of an SSD.  In this evaluation, we focus on examining how the
durability of an SSD is affected by different GC algorithms.

To study the durability of an SSD, we have to make the SSD continue
handling a sufficient number of I/O requests until it is worn out.
In order to speed up our simulation, we decrease the maximum number
of erasures sustainable by each block to 50.  We also reduce the
size of the SSD such that each flash package contains 4096 blocks,
so the  size of each flash package is 1GB.  Other configurations are
the same as we described in \S\ref{subsec: configuration}.  Also,
instead of using the real-world traces as in previous simulations,
we drive the simulation with the synthetic traces that have more
aggressive I/O rates so that the SSD is worn out soon. Specifically,
we consider the same set of synthetic traces {\sf Random},  {\sf
Hybrid} and {\sf Sequential} as described in \S\ref{subsec:
validation on fixed-point}, but here we set the mean inter-arrival
time of I/O requests to be 10ms (as opposed to 100ms in
\S\ref{subsec: validation on fixed-point}) based on Poisson
arrivals.


Due to the use of bad block management \cite{micron11}, an SSD can allow a
small percentage of bad (worn-out) blocks during its lifetime.  Suppose that
the SSD can allow up to $e$\% of bad blocks for some parameter $e$. To derive
the durability of the SSD,  we first continue running each workload trace on
the SSD until $e$\% blocks are worn out, i.e.,  the erasure limit is reached.
Then we record the length of the duration span that the SSD survives, and take
it as the durability of the SSD.
For comparison, we normalize the durability with respect to that of the greedy
algorithm (which is expected to have the minimum durability). In this
experiment, we consider the case where $e$\% = 5\%, while we also verify that
similar observations are made for other values of $e\% \le 10\%$.  Also, we
assume that the SSD is brand-new (i.e., the initial state is empty) and all
blocks have no erasure at the beginning.

\begin{figure}[!htb]
  \centering
  \includegraphics[width=\linewidth]{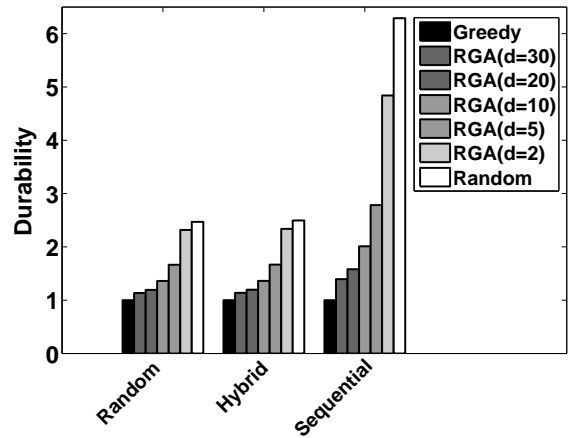}\label{fig: empty_durability5}
  \caption{Durability of different GC algorithms (normalized with respect to
  the greedy algorithm).}
  \label{fig: durability}
\end{figure}

Figure~\ref{fig: durability} shows the results.  We observe that the
durability results of different GC algorithms are consistent with those of
wear-leveling in Figure~\ref{fig: wear_leveling}.  We observe that the random
algorithm achieves the maximum durability, and the value can be almost six
times over that of the greedy algorithm (e.g., in the {\sf Sequential}
workload).  Again, RGA provides a tunable durability between the random and
greedy algorithms.  When the window size $d\le 5$, the durability of RGA can
be within 68\% of the maximum lifetime of the random algorithm for
{\sf Random} and {\sf Hybrid} workloads. For the {\sf Sequential} workload,
the durability of RGA drops to 40\% of the maximum lifetime of the random
algorithm when $d=5$. However, it is still almost 3 times higher than that of
the greedy algorithm.

\subsection{Summary}

From the above simulations, we see that the greedy algorithm performs the best
and the random algorithm performs the worst in terms of cleaning cost and I/O
throughput, while the opposite holds in terms of wear-leveling and durability.
We demonstrate that our RGA provides a tradeoff spectrum between the two
algorithms by tuning the window size.  This simulation study not only confirms
our theoretical model, but also shows that our RGA can be viewed as an
effective tunable algorithm to balance between throughput performance and
durability of an SSD.

\section{Related Work}
\label{sec: related}

The research on NAND-flash based SSDs has recently received a lot of
attention.  Many aspects of SSDs are being studied. A survey on
algorithms and data structures for flash memories can be found in
\cite{Gal05}. Kawaguchi et al. \cite{Kawaguchi95} propose a
flash-based file system based on the log-structured file system
design. Birrell et al. \cite{Birrell07} propose new data structures
to improve the write performance of SSDs, and Gupta et al.
\cite{Gupta11} suggest to exploit value locality and design content
addressable SSDs so as  to optimize the performance.  Matthews et
al. \cite{Matthews08} use NAND-based disk caching to mitigate the
I/O bottlenecks of HDDs, and Kim et al. \cite{Kim11} consider hybrid
storage by combining SSDs and HDDs. Agrawal et al. \cite{Agrawal08}
study different design tradeoffs of SSDs via a trace-driven
simulator based on DiskSim \cite{disksim}. Chen et al. \cite{Chen09}
further reveal many intrinsic characteristics of SSDs via empirical
measurements. Polte et al. \cite{Polte09} also study the performance
of SSDs via experiments, and Park et al. \cite{Park11} mainly focus
on the energy efficiency of SSDs.  Note that \cite{Agrawal08}
addresses the tradeoff between cleaning cost and wear-leveling in
GC, but it is mainly based on empirical evaluation.

A variety of wear-leveling techniques have been proposed, mainly from
an applied perspective.  Some of them are proposed in patents
\cite{wear-leveling_patent1, wear-leveling_patent2, wear-leveling_patent3,
wear-leveling_patent4, wear-leveling_patent5, wear-leveling_patent6,
wear-leveling_patent7}. Several research papers have been proposed to maximize
wear-leveling in SSDs based on {\em hot-cold swapping}, whose main idea is to
swap the frequently-used hot data in worn blocks and the rarely-used cold data
in new blocks.  For example, Chiang et al.  \cite{Chiang99a, Chiang99b}
propose clustering methods for hot/cold data based on access patterns to
maximize wear-leveling.  Jung et al. \cite{Jung07} propose a memory-efficient
design for wear-leveling by tracking only block groups, while maintaining
wear-leveling performance.  Authors of \cite{Chang09,Chang10,Chang11} also
propose different strategies based on hot-cold swapping to further improve the
wear-leveling performance.  Our work differs from above studies in that we
focus on characterizing the optimal tradeoff of GC algorithms, such that we
provide flexibility for SSD practitioners to reduce wear-leveling to trade for
higher cleaning performance.  We also propose a tunable GC algorithm to
realize the tradeoff.


From a theoretical perspective, some studies propose analytical
frameworks to quantify the performance of GC algorithms.  A
comparative study between online and offline wear-leveling policies
is presented in \cite{benaroya06}. Hu et al. \cite{Hu09} propose a
probabilistic model to quantify the additional writes due to GC
(i.e., the cleaning cost defined in our work). They study a modified
greedy GC algorithm, and implement an event-driven simulator to
validate their model. Bux and Iliadis \cite{Bux10} propose
theoretical models to analyze the greedy GC algorithm under the
uniform workload, and Desnoyers \cite{Desnoyers12} also analyzes the
performance of LRU and greedy GC algorithms when page-level address
mapping is used. Our work differs from them  in the following.
First, the previous work focuses on analyzing the write
amplification which corresponds to the cleaning cost in our paper,
but our focus is to analyze the tradeoff between cleaning cost and
wear-leveling, which are both very important in designing GC
algorithms, and further explore the design space of GC algorithms.
Second, our analytical models are also very different. In
particular, we use a Markov model to characterize the I/O dynamics
of SSDs and adapt the mean field technique to approximate
large-scale systems, then we  develop an optimization framework to
derive the optimal tradeoff curve. Finally, our model also provides
a good approximation under general workload and address mapping, and
it is further validated via trace-driven evaluation.

We note that an independent analytical work \cite{vanhoudt13}, which is
published in the same conference as ours, also applies the mean-field
technique to analyze different GC algorithms. Its $d$-choices GC algorithm has
the same construction as our RGA.  Our work has the following key differences.
First, similar to prior analytical studies, the work \cite{vanhoudt13}
focuses on write amplification, while we focus on the trade-off between
cleaning cost and wear leveling.  Second, its analysis is limited to the
uniform workload only, while we also address the general workload.  Finally,
we validate our analysis via trace-driven simulations, which are not
considered in the work \cite{vanhoudt13}. 


\section{Conclusions}
\label{sec: conclusion}

In this paper, we propose an analytical model to characterize the
performance-durability tradeoff of an SSD.  We model the I/O dynamics of a
large-scale SSD, and use the mean field theory to derive the asymptotic
results in equilibrium.  In particular, we classify the blocks of an SSD into
different types according to the number of valid pages contained in each
block, and our mean field results can provide effective approximation on the
fraction of different types of blocks in the steady state even under the
general workload.  We define two metrics, namely cleaning cost and
wear-leveling, to quantify the performance of GC algorithms.  In particular,
we theoretically characterize the optimal tradeoff curve between cleaning cost
and wear-leveling, and develop an optimization framework to explore the full
optimal design space of GC algorithms.  Inspired from our analytical
framework, we develop a tunable GC algorithm called the randomized greedy
algorithm (RGA) which can efficiently balance the tradeoff between cleaning
cost and wear-leveling by tuning the parameter of the window size $d$.  We use
trace-driven simulation based on DiskSim with SSD add-ons to validate our
analytical model, and show the effectiveness of RGA in tuning the
performance-durability tradeoff in deployment.

\bibliographystyle{abbrv}
\bibliography{pl}

\begin{thebibliography}{10}

\bibitem{Agrawal08}
N.~Agrawal, V.~Prabhakaran, T.~Wobber, J.~D. Davis, M.~Manasse, and
  R.~Panigrahy.
\newblock {Design Tradeoffs for SSD Performance}.
\newblock In {\em Proc. of USENIX ATC}, Jun 2008.

\bibitem{wear-leveling_patent1}
M.~Assar, S.~Nemazie, and P.~Estakhri.
\newblock {Flash Memorymass Storage Architecture Incorporation Wear Leveling
  Technique}.
\newblock US patent 5,479,638, Dec 1995.

\bibitem{wear-leveling_patent7}
A.~Ban.
\newblock {Wear Leveling of Static Areas in Flash Memory}.
\newblock US patent 6732221, May 2004.

\bibitem{benaroya06}
A.~Ben-Aroya and S.~Toledo.
\newblock {Competitive Analysis of Flash-memory Algorithms}.
\newblock In {\em Proc. of Annual European Symposium}, Sep 2006.

\bibitem{meanfield_boundec}
M.~Bena{\"{\i}}m and J.-Y.~L. Boudec.
\newblock {A Class of Mean Field Interaction Models for Computer and
  Communication Systems}.
\newblock {\em Performance Evaluation}, 2008.

\bibitem{Birrell07}
A.~Birrell, M.~Isard, C.~Thacker, and T.~Wobber.
\newblock {A Design for High-performance Flash Disks}.
\newblock {\em ACM SIGOPS Oper. Syst. Rev.}, 41(2):88--93, Apr 2007.

\bibitem{wear-leveling_patent2}
R.~H. Bruce, R.~H. Bruce, E.~T. Cohen, and A.~J. Christie.
\newblock {Unified Re-map and Cache-index Table with Dual Write-counters for
  Wear-leveling of Non-volitile Flash Ram Mass Storage}.
\newblock US patent 6,000,006, Dec 1999.

\bibitem{disksim}
J.~S. Bucy, J.~Schindler, S.~W. Schlosser, and G.~R. Ganger.
\newblock {The DiskSim Simulation Environment Version 4.0 Reference Manual}.
\newblock Technical Report CMUPDL-08-101, Carnegie Mellon University, May 2008.

\bibitem{Bux10}
W.~Bux and I.~Iliadis.
\newblock {Performance of Greedy Garbage Collection in Flash-based Solid-state
  Drives}.
\newblock {\em Performance Evaluation}, November 2010.

\bibitem{Chang09}
L.-P. Chang and C.-D. Du.
\newblock {Design and Implementation of an Efficient Wear-leveling Algorithm
  for Solid-state-disk Microcontrollers}.
\newblock {\em ACM Trans. Des. Autom. Electron. Syst.}, 15(1):6:1--6:36, Dec
  2009.

\bibitem{Chang11}
L.-P. Chang and L.-C. Huang.
\newblock {A Low-cost Wear-leveling Algorithm for Block-mapping Solid-state
  Disks}.
\newblock In {\em Proc of SIGPLAN/SIGBED Conf. on LCTES}, Apr 2011.

\bibitem{Chang10}
Y.-H. Chang, J.-W. Hsieh, and T.-W. Kuo.
\newblock {Improving Flash Wear-Leveling by Proactively Moving Static Data}.
\newblock {\em IEEE Tran. on Computers}, 59:53--65, Jan 2010.

\bibitem{Chen09}
F.~Chen, D.~A. Koufaty, and X.~Zhang.
\newblock {Understanding Intrinsic Characteristics and System Implications of
  Flash Memory Based Solid State Drives}.
\newblock In {\em Proc. of ACM SIGMETRICS}, Jun 2009.

\bibitem{Chiang99b}
M.-L. Chiang and R.-C. Chang.
\newblock {Cleaning Policies in Mobile Computers Using Flash Memory}.
\newblock {\em J. Syst. Softw.}, 48(3):213--231, Nov 1999.

\bibitem{Chiang99a}
M.-L. Chiang, P.~C.~H. Lee, and R.-C. Chang.
\newblock {Using Data Clustering to Improve Cleaning Performance for Flash
  Memory}.
\newblock {\em Softw. Pract. Exper.}, 29(3):267--290, Mar 1999.

\bibitem{Chung06}
T.-S. Chung, D.-J. Park, S.~Park, D.-H. Lee, S.-W. Lee, and H.-J. Song.
\newblock {System Software For Flash Memory: A Survey}.
\newblock In {\em Proc. of Int. Conf. on Embedded and Ubiquitous Computing},
  2006.

\bibitem{Chung09}
T.-S. Chung, D.-J. Park, S.~Park, D.-H. Lee, S.-W. Lee, and H.-J. Song.
\newblock {A Survey of Flash Translation Layer}.
\newblock {\em Journal of Systems Architecture}, 55(5-6):332--343, May 2009.

\bibitem{Desnoyers12}
P.~Desnoyers.
\newblock {Analytic Modeling of SSD Write Performance}.
\newblock In {\em Proceedings of SYSTOR}, 2012.

\bibitem{SSDinDataCenter2}
R.~Enderle.
\newblock {Revolution in January: EMC Brings Flash Drives into the Data
  Center}.
\newblock \url{http://www.itbusinessedge.com/blogs/rob/?p=184}, Jan 2008.

\bibitem{wear-leveling_patent3}
P.~Estakhri, M.~Assar, R.~Reid, Alan, and B.~Iman.
\newblock {Method of and Architecture for Controlling System Data with
  Automatic Wear Leveling in a Semiconductor Non-volitile Mass Storage Memory}.
\newblock US patent 5,835,935, Nov 1998.

\bibitem{SSDPriceDecrease}
D.~Floyer.
\newblock {Flash Pricing Trends Disrupt Storage}.
\newblock \url{http://wikibon.org/wiki/v/Flash_Pricing_Trends_Disrupt_Storage},
  May 2010.

\bibitem{Gal05}
E.~Gal and S.~Toledo.
\newblock {Algorithms and Data Structures for Flash Memories}.
\newblock {\em ACM Computing Surveys}, 37(2):138--163, Jun 2005.

\bibitem{FAST12bleakfeature}
L.~M. Grupp, J.~D. Davis, and S.~Swanson.
\newblock {The Bleak Future of NAND Flash Memory}.
\newblock In {\em Proc. of USENIX FAST}, 2012.

\bibitem{Gupta09}
A.~Gupta, Y.~Kim, and B.~Urgaonkar.
\newblock {DFTL: A Flash Translation Layer Employing Demand-based Selective
  Caching of Page-level Address Mappings}.
\newblock In {\em Proc. of ACM ASPLOS}, March 2009.

\bibitem{Gupta11}
A.~Gupta, R.~Pisolkar, B.~Urgaonkar, and A.~Sivasubramaniam.
\newblock {Leveraging Value Locality in Optimizing NAND Flash-based SSDs}.
\newblock In {\em Proc. of USENIX FAST}, 2011.

\bibitem{wear-leveling_patent4}
S.-W. Han.
\newblock {Flash Memory Wear Leveling System and Method}.
\newblock US patent 6,016,275, Jan 2000.

\bibitem{SSDinDataCenter3}
K.~Hess.
\newblock {2011: Year of the SSD?}
\newblock
  \url{http://www.datacenterknowledge.com/archives/2011/02/17/2011-year-of-the-ssd/},
  Feb 2011.

\bibitem{Hu09}
X.-Y. Hu, E.~Eleftheriou, R.~Haas, I.~Iliadis, and R.~Pletka.
\newblock {Write Amplification Analysis in Flash-based Solid State Drives}.
\newblock In {\em Proc. of SYSTOR}, May 2009.

\bibitem{fainess_index}
R.~K. Jain, D.-M.~W. Chiu, and W.~R. Hawe.
\newblock {A Quantitative Measure Of Fairness And Discrimination For Resource
  Allocation In Shared Computer Systems}.
\newblock Technical report, DEC, 1984.

\bibitem{Jung07}
D.~Jung, Y.-H. Chae, H.~Jo, J.-S. Kim, and J.~Lee.
\newblock {A Group-based Wear-leveling Algorithm for Large-capacity Flash
  Memory Storage Systems}.
\newblock In {\em Proc. of Int. Conf. on Compilers, Architecture, and Synthesis
  for Embedded Systems}, Sep 2007.

\bibitem{Kawaguchi95}
A.~Kawaguchi, S.~Nishioka, and H.~Motoda.
\newblock {A Flash-memory Based File System}.
\newblock In {\em Proc. of USENIX Technical Conference}, Jan 1995.

\bibitem{Kim11}
Y.~Kim, A.~Gupta, B.~Urgaonkar, P.~Berman, and A.~Sivasubramaniam.
\newblock {HybridStore: A Cost-Efficient, High-Performance Storage System
  Combining SSDs and HDDs}.
\newblock In {\em Proc. of IEEE MASCOTS}, Jul 2011.

\bibitem{Lee07}
S.-W. Lee, D.-J. Park, T.-S. Chung, D.-H. Lee, S.~Park, and H.-J. Song.
\newblock {A Log Buffer-based Flash Translation Layer Using Fully-associative
  Sector Translation}.
\newblock {\em ACM Trans. on Embedded Computing Systems}, 6(3), Jul 2007.

\bibitem{wear-leveling_patent5}
K.~M.~J. Lofgren, R.~D. Norman, G.~B. Thelin, and A.~Gupta.
\newblock {Wear Leveling Techniques for Flash EEPROM Systems}.
\newblock US patent 6,850,443, Feb 2005.

\bibitem{Matthews08}
J.~Matthews, S.~Trika, D.~Hensgen, R.~Coulson, and K.~Grimsrud.
\newblock {Intel$^{\circledR}$ Turbo Memory: Nonvolatile Disk Caches in the
  Storage Hierarchy of Mainstream Computer Systems}.
\newblock {\em ACM Trans. on Storage}, 4(2):4:1--4:24, May 2008.

\bibitem{micron11}
{Micron Technology}.
\newblock {Bad Block Management in NAND Flash Memory}.
\newblock Technical Note, TN-29-59, 2011.

\bibitem{meanfield_loadbalance}
M.~Mitzenmacher.
\newblock {Load Balancing and Density Dependent Jump Markov Processes}.
\newblock In {\em Proc. of IEEE FOCS}, Oct. 1996.

\bibitem{MSST11replaytrace}
M.~Murugan and D.~Du.
\newblock {Rejuvenator: A Static Wear Leveling Algorithm for NAND Flash Memory
  with Minimized Overhead}.
\newblock In {\em Proc. of IEEE MSST}, May 2011.

\bibitem{Park08}
C.~Park, W.~Cheon, J.~Kang, K.~Roh, W.~Cho, and J.-S. Kim.
\newblock {A Reconfigurable FTL (Flash Translation Layer) Architecture for NAND
  Flash-based Applications}.
\newblock {\em ACM Trans. Embed. Comput. Syst.}, 7(4):38:1--38:23, Aug 2008.

\bibitem{Park11}
S.~Park, Y.~Kim, B.~Urgaonkar, J.~Lee, and E.~Seo.
\newblock {A Comprehensive Study of Energy Efficiency and Performance of
  Flash-based SSD}.
\newblock {\em Journal of Systems Architecture}, 57(4):354--365, April 2011.

\bibitem{Polte09}
M.~Polte, J.~Simsa, and G.~Gibson.
\newblock {Enabling Enterprise Solid State Disks Performance}.
\newblock In {\em 1st Workshop on Integrating Solid-state Memory into the
  Storage Hierarchy}, March 2009.

\bibitem{Qin10}
Z.~Qin, Y.~Wang, D.~Liu, and Z.~Shao.
\newblock {Demand-based Block-level Address Mapping in Large-scale NAND Flash
  Storage Systems}.
\newblock In {\em Proc. of IEEE/ACM/IFIP CODES+ISSS}, Oct 2010.

\bibitem{umasstrace}
{Storage Performance Council}.
\newblock \url{http://traces.cs.umass.edu/index.php/Storage/Storage}, 2002.

\bibitem{vanhoudt13}
B.~{Van Houdt}.
\newblock {A Mean Field Model for a Class of Garbage Collection Algorithms in
  Flash-based Solid State Drives}.
\newblock In {\em Proc. of ACM SIGMETRICS}, 2013.

\bibitem{fiutrace}
A.~Verma, R.~Koller, L.~Useche, and R.~Rangaswami.
\newblock {SRCMap: Energy Proportional Storage using Dynamic Consolidation}.
\newblock In {\em Proc. of USENIX FAST}, Feb 2010.
\newblock \url{http://sylab.cs.fiu.edu/projects/srcmap/start}.

\bibitem{wear-leveling_patent6}
S.~E. Wells.
\newblock {Method for Wear Leveling in a Flash EEPROM Memory}.
\newblock US patent 5,341,339, Aug 1994.

\end{thebibliography}

\newpage
\appendix

\section{Proof of Convergence}

We now  formally prove the convergence of the SSD system state under the
uniform workload.  Our proof consists of two parts.  We first prove that the
stochastic process $\boldsymbol{M}^N(t)$ indeed converges to the
deterministic process $\boldsymbol{s}(t)$ when $N \rightarrow \infty$.  We
then prove that the deterministic process described in Equation~(\ref{eq:
ODE})  converges  to the unique fixed point $\boldsymbol{\pi}$ in
Equation~(\ref{eq: fixed point}).

We first show that the re-scaled process $\widetilde{\boldsymbol{M}}^N(t)$
converges to $\boldsymbol{s}(t)$. Let us first show several important
properties of the stochastic process $\boldsymbol{M}^N(t)$.

\begin{lemma}
Define $S=\{\boldsymbol{m} \! \in \!  R^{k+1} | \sum_{i=0}^k
m_i\!=\!1, m_i \geq 0 \, \forall i\}$. For any $\boldsymbol{s} \in
S$, let $\boldsymbol{f}^N(\boldsymbol{s})
= E(\boldsymbol{M}^N(t+1)-\boldsymbol{M}^N(t) | \boldsymbol{M}^N(t)
=\boldsymbol{s})$ be the expected change to the occupancy measure in one time
slot, and let $\varepsilon(N)=1/N$. We have
$\lim_{N\rightarrow \infty}\varepsilon(N)$$ = 0$, and
$\lim_{N\rightarrow \infty}
\frac{\boldsymbol{f}^N(\boldsymbol{s})}{\varepsilon(N)}$ exists for
 $\forall \boldsymbol{s} \in S$. \label{lem: C1}
\end{lemma}

\noindent{\bf Proof: } Since
$\varepsilon(N)=\frac{1}{N}$, $\lim_{N\rightarrow
\infty}\varepsilon(N) = 0$. Denote
$\boldsymbol{f}^N(\boldsymbol{s})=(f^N_0(\boldsymbol{s}),
f^N_1(\boldsymbol{s}), ..., f^N_k(\boldsymbol{s}))$. Consider the
expected change in $M_i(t)$ ($1 \leq i \leq k-1$) during one time
slot when $\lambda$ requests arrive. For each request, the
probability of changing a block of type $i$ to other states is
$\frac{1}{N}Ns_i$, and the corresponding change in  $M_i(t)$ is
$-\frac{1}{N}$. A request may also change blocks of type $i+1$ (or type $i-1$)
to state $i$, with probability being $\frac{i+1}{Nk}Ns_{i+1}$ (or
$\frac{k-i+1}{Nk}Ns_{i-1}$) and the change in $M_i(t)$ being
$\frac{1}{N}$ (or $\frac{1}{N}$). Thus, we have
\begin{equation*}
  f^N_i(\boldsymbol{s} )\!=\! - \frac{1}{N}\lambda s_i
+ \frac{i+1}{N k}\lambda s_{i+1} + \frac{k-i+1}{N k}\lambda s_{i-1}, 1\! \leq\! i\! \leq k\!-\!1.
\end{equation*}
Similarly, we can  also derive $f^N_0(\boldsymbol{s})$ and
$f^N_k(\boldsymbol{s})$.
Therefore, we have $\lim_{N\rightarrow \infty}
   \frac{\boldsymbol{f}^N(\boldsymbol{s})}{\varepsilon(N)}
  = \boldsymbol{f}(\boldsymbol{s})$, where
\begin{equation*}
    \left\{
    \begin{aligned}
     f_i(\boldsymbol{s}) =& - \lambda s_i \!+\! \frac{i+1}{ k}\lambda s_{i+1}
        \!+\! \frac{k-i+1}{k}\lambda s_{i-1}, \quad 1 \!\leq\! i \!\leq\! k\!-\!1 ,\\
     f_0(\boldsymbol{s}) =& - \lambda s_0 + \frac{1}{ k}\lambda s_{1};\,\,\,
     f_k(\boldsymbol{s}) = - \lambda s_k + \,\frac{1}{ k}\lambda s_{k-1}.
     \hspace*{28pt} \rule{1.8mm}{2.5mm}
    \end{aligned}
    \right.
\end{equation*}

\begin{lemma}
Define $W^N(t)$ as an upper bound on the number of blocks that make
a transition in time slot $t$. Then $W^N(t)$ satisfies
$E(W^N(t)^2|\boldsymbol{M}^N(t)=\boldsymbol{s}) \leq
cN^2\varepsilon(N)^2$ where $c$ is a constant.
\label{lem: C2}
\end{lemma}
 \noindent{\bf Proof: } During the time slot $t$,
$\lambda$ requests arrive, and for each request, it accesses a block
with probability $\frac{1}{N}$. Therefore, $W^N(t)$ follows a
binomial distribution with parameters $\frac{\lambda}{N}$ and $N$.
\begin{equation*}
    E(W^N(t)^2|\boldsymbol{M}^N(t)\!=\!\boldsymbol{s}) \!=\!
\left(\!N \frac{\lambda}{N}\!\right)^2 \!+\!
   N \frac{\lambda}{N}\left(\!1\!-\!\frac{\lambda}{N}\!\right)
   \leq \lambda^2 \!+\! \lambda,
\end{equation*}
which shows the result in  Lemma \ref{lem: C2} with $c= \lambda^2 +
\lambda$. \done

\begin{lemma}
There exists $\beta > 0$ and a function $\varphi(\boldsymbol{s},
\alpha)$ defined on $S\times[0, \beta]$ such that $\varphi$ has
continuous derivatives everywhere and
$\frac{\boldsymbol{f}^N(\boldsymbol{s})}
{\varepsilon(N)}=\varphi(\boldsymbol{s},\frac{1}{N})$.
  \label{lem: C3}
\end{lemma}
\noindent{\bf Proof: } From the proof of Lemma \ref{lem: C1},
$\frac{\boldsymbol{f}^N(\boldsymbol{s})}{\varepsilon(N)}
 =\boldsymbol{f}(\boldsymbol{s})$.
Since $\boldsymbol{f}(\boldsymbol{s})$ is a rational function with
respect to $\boldsymbol{s}$ and $\frac{1}{N}$, \mbox{Lemma \ref{lem:
C3}}  holds. \done

Now, we can now show that the re-scaled process
$\widetilde{\boldsymbol{M}}^N(t)$ converges to $\boldsymbol{s}(t)$, with the
following theorem.

\begin{theorem}
If $\boldsymbol{M}^N(0) \rightarrow \boldsymbol{m}$ in probability
as $N\rightarrow \infty$, then for all $T > 0$, $\sup_{0\leq t \leq
T}\| \widetilde{\boldsymbol{M}}^N(t) - \boldsymbol{s}(t)\|
\rightarrow 0$ in probability, where $\boldsymbol{s}(t)$ satisfies
the ODEs in Equation (\ref{eq: ODE}) and $\boldsymbol{s}(0) =
\boldsymbol{m}$.
  \label{theo: convergence to ODE}
\end{theorem}
 \noindent {\bf Proof: } The theorem holds due to
Lemmas~\ref{lem: C1}, \ref{lem: C2}, and \ref{lem: C3}, and the existing
theorem in \cite{meanfield_boundec} (Corollary 1).  \done

 Note that the theorem in \cite{meanfield_boundec} provides a
way to prove the convergence to mean field limit, provided that several
sufficient conditions hold. Therefore, to invoke the theorem in
\cite{meanfield_boundec}, we must explicitly verify that our model indeed
satisfies the conditions, i.e., Lemma \ref{lem: C1}-\ref{lem: C3}, so as to
make the proof complete.


\begin{corollary}
If $\boldsymbol{M}^N(0) \rightarrow \boldsymbol{m}$ in probability
as $N\rightarrow \infty$, then for all $T > 0$, $\sup_{0\leq t \leq
T}\| \boldsymbol{M}^N(t) - \boldsymbol{s}(\frac{t}{N})\| \rightarrow
0$ in probability, where $\boldsymbol{s}(t)$ satisfies the ODEs in
Equation (\ref{eq: ODE}) and $\boldsymbol{s}(0) = \boldsymbol{m}$.
\label{corollary: convergence to ODE}
\end{corollary}

In the following, we prove that the deterministic process
$\boldsymbol{s}(t)$ in Equation (\ref{eq: ODE}) converges  to the unique
fixed point $\boldsymbol{\pi}$ in Equation (\ref{eq: fixed point}). Note
that the ODEs in Equation~(\ref{eq: ODE}) is just a special case of the ODEs
in Equation~(\ref{eq: ODE_general}). Since our proof also applies to the
general case, to avoid redundancy, we directly present the convergence proof
of the general case, i.e., the convergence of the ODEs in Equation~(\ref{eq:
ODE_general}) to the fixed point in Equation~(\ref{eq:
fixed_point_general}).
The detailed proof is shown in Theorem \ref{theorem: convergence to fixed
point general}. We thank professor Benny Van Houdt for giving us invaluable
comments on this proof.

\begin{theorem}
The deterministic process $\boldsymbol{s}(t)$ which is specified by ODEs
(\ref{eq: ODE_general}) converges to the fixed point $\boldsymbol{\pi}$
which is determined by Equation (\ref{eq: fixed_point_general}).
  \label{theorem: convergence to fixed point general}
\end{theorem}
 \noindent {\bf Proof: } Note that Equation~(\ref{eq: ODE_general}) can be
 rewritten as follows.
 \begin{equation}
 \frac{d\bs(t)}{dt}=\bs(t)\bQ,
 \label{eq:ODE_general_matrix}
 \end{equation}
 where $\bs(t)=(s_0(t),s_1(t),...,s_k(t))$ and $\bQ=[q_{i,j}]$.
\begin{equation*}
    q_{i,j}=\left\{
    \begin{aligned}
      &\lambda p_{i,j}, &\mbox{for} j\neq i,\\
      &-\lambda p_{0,1}, &\mbox{ for } j=i=0,\\
      &-\lambda p_{k,k-1}, &\mbox{ for } j=i=k,\\
      &-\lambda (p_{i,i-1}+p_{i,i+1}), &\mbox{ for } 0<j=i<k.
    \end{aligned}\right.
\end{equation*}
Note that if we treat the state transition of a particular block shown in
Figure~\ref{fig: state transition} as a birth-death process, then
Equation~(\ref{eq:ODE_general_matrix}) exactly maps to the Kolmogorov's
forward equations where $\bQ$ is just the rate matrix of the birth-death
process. Therefore, $\bs(t)$ converges to the stationary distribution of the
birth-death process $\boldmath{\pi}$ where $\boldmath{\pi}\bQ=\boldmath{0}$.
We can easily verity that the fixed point $\boldmath{\pi}$ in
Equation~(\ref{eq: fixed_point_general}) satisfies the condition
$\boldmath{\pi} \bQ=\boldmath{0}$, which completes the proof. \done

Note that Theorem \ref{theorem: convergence to fixed point general} also
completes the proof that the ODEs in  Equation (\ref{eq: ODE}) converges  to
the unique fixed point $\boldsymbol{\pi}$ in Equation (\ref{eq: fixed
point}).

\section{Proof of Optimal Design Space}
	
We now prove Theorem~\ref{theo: design_space} in \S\ref{subsec: exploring}.
We solve Equation~(\ref{problem: F VS C_max}) by minimizing the inverse of the
objective function, and the problem is a convex optimization problem.
If a point $(\boldsymbol{\tilde{w}},
\boldsymbol{\tilde{u}}, \tilde{v_1}, \tilde{v_2})$ satisfies the KKT
conditions which are stated in Equation (\ref{eq: KKT}), then
$\boldsymbol{\tilde{w}}$ is the global minimum.
\begin{equation}
\left\{\begin{aligned}
  &2 w_i\pi_i - u_i + v_1\pi_i + v_2 i \pi_i = 0;\,\,\, u_i \geq 0;  \,\,\, w_i \geq 0;\\
  &u_iw_i = 0;\,\,\,\sum\nolimits_{i=0}^k w_i \pi_i = 1;\,\,\, \sum\nolimits_{i=0}^k i w_i\pi_i = \mathcal{C^*}.
\end{aligned}
\right.
\label{eq: KKT}
\end{equation}

To find a point satisfying the KKT conditions, we first consider the case
when $0<\mathcal{C^*}< \sum_{i=0}^{k}i\pi_i$. Let
\begin{equation}
\AI_1 = \min_{0\leq j\leq k}\{j: \sum\nolimits_{i=0}^j i\pi_i \!-\! \mathcal{C^*}\sum\nolimits_{i=0}^j \pi_i > 0\}.
  \label{eq: step1}
\end{equation}
 Note that $\AI_1$ must exist because $\mathcal{C^*}\!<\! \sum_{i=0}^{k}i\pi_i$ and
$\sum_{i=0}^{k}\pi_i\!=\!1$. Clearly, we  have $\AI_1 >
\mathcal{C^*}$ and $\sum_{i=0}^j i\pi_i \!-\!
\mathcal{C^*}\sum_{i=0}^j\pi_i > 0\, \mbox{ for }\AI_1\leq j \leq
k.$ Moreover, we  have $\sum_{i=0}^j i^2\pi_i -
\mathcal{C^*}\sum_{i=0}^ji\pi_i > 0$ ($\AI_1\leq j \leq k$). Now we
prove that the following inequality holds.
\begin{equation}
\frac{\sum_{i=0}^{\AI_1} i^2\pi_i -
   \mathcal{C^*}\sum_{i=0}^{\AI_1} i \pi_i}{\sum_{i=0}^{\AI_1} i\pi_i -
    \mathcal{C^*}\sum_{i=0}^{\AI_1} \pi_i} > \AI_1.
\label{eq: step2}
\end{equation}
To prove the inequality (\ref{eq: step2}), we rewrite the left hand
side of the inequality as follows.
\begin{equation*}
  \begin{aligned}
    &\frac{\sum_{i=0}^{\AI_1} i^2\pi_i -
   \mathcal{C^*}\sum_{i=0}^{\AI_1} i \pi_i}{\sum_{i=0}^{\AI_1} i\pi_i -
    \mathcal{C^*}\sum_{i=0}^{\AI_1} \pi_i}
    \triangleq&\frac{-ax+by+\AI_1z}{-x+y+z},
  \end{aligned}
\end{equation*}
where $x=-\sum_{i=0}^{\lfloor\mathcal{C^*}\rfloor}
(i-\mathcal{C^*})\pi_i$,
$y=\sum_{i=\lfloor\mathcal{C^*}\rfloor+1}^{\AI_1-1}
(i-\mathcal{C^*})\pi_i$ and $z=(\AI_1-\mathcal{C^*})\pi_{\AI_1}$.
Clearly, we have $x>0, y\geq0, z>0$ and $\AI_1>b>a>0$. Since $\AI_1$
is the smallest integer which satisfies the condition in (\ref{eq:
step1}), we also have $-x+y<0$ and $-x+y+z>0$. Now, if $-ax+by\geq
0$, then inequality (\ref{eq: step2}) holds.   Otherwise,
\begin{equation*}
  \begin{aligned}
    & \frac{-ax+by+\AI_1z}{-x+y+z}
    = \AI_1+\frac{(\AI_1-a)(x-y)+(b-a)y}{-x+y+z}\\
    > &\AI_1 (\mbox{ as } \AI_1>b>a>0, -x+y<0, \mbox{ and } -x+y+z>0 ).
  \end{aligned}
\end{equation*}

Now, we argue that there exists an $\AI$ ($\AI_1\leq \AI\leq k$)
such that
\begin{equation}
\left\{\begin{aligned}
  &\AI\!<\!k \mbox{ and } \AI\! <\! \frac{\sum_{i=0}^{\AI} i^2\pi_i -
   \mathcal{C^*}\sum_{i=0}^{\AI} i \pi_i}{\sum_{i=0}^{\AI} i\pi_i -
    \mathcal{C^*}\sum_{i=0}^{\AI} \pi_i} \!\leq \!\AI+1, \mbox{ or }\\
  &\AI \!=\! k \mbox{ and } \AI \!< \!\frac{\sum_{i=0}^{\AI} i^2\pi_i -
   \mathcal{C^*}\sum_{i=0}^{\AI} i \pi_i}{\sum_{i=0}^{\AI} i\pi_i -
    \mathcal{C^*}\sum_{i=0}^{\AI} \pi_i}.
\end{aligned}
\right.
\label{eq: step3}
\end{equation}
To prove it, we can examine from $\AI_1$. Since inequality (\ref{eq:
step2}) holds, if $\AI_1 < k$ and $\frac{\sum_{i=0}^{\AI_1} i^2\pi_i
- \mathcal{C^*}\sum_{i=0}^{\AI_1} i \pi_i}{\sum_{i=0}^{\AI_1} i\pi_i
- \mathcal{C^*}\sum_{i=0}^{\AI_1} \pi_i} > \AI_1+1$, then we have
$$\frac{\sum_{i=0}^{\AI_1+1} i^2\pi_i -
\mathcal{C^*}\sum_{i=0}^{\AI_1+1} i \pi_i}{\sum_{i=0}^{\AI_1+1}
i\pi_i - \mathcal{C^*}\sum_{i=0}^{\AI_1+1} \pi_i} > \AI_1+1.$$
Therefore, either we find an $\AI$ such that $\AI \!\! <\!\!
\frac{\sum_{i=0}^{\AI} \! i^2\pi_i - \mathcal{C^*}\sum_{i=0}^{\AI}
\! i \pi_i}{\sum_{i=0}^{\AI} \! i\pi_i -
    \mathcal{C^*}\sum_{i=0}^{\AI} \! \pi_i}\!\! \leq \!\!\AI+1$ or we reach $k$.
Now, given the $\AI$ in Equation~(\ref{eq: step3}), we define
\begin{equation*}
\left\{\begin{aligned}
  X_{\AI}&\!\!=\!\!\sum\nolimits_{i=0}^\AI\! i^2\pi_i \!-\! \mathcal{C^*}\!\sum\nolimits_{i=0}^\AI\! i \pi_i,\,\,
  Y_{\AI}\!\!=\!\!\sum\nolimits_{i=0}^\AI\! i\pi_i \!-\! \mathcal{C^*}\!\sum\nolimits_{i=0}^\AI\! \pi_i,\\
  Z_{\AI}&=\sum\nolimits_{i=0}^\AI \pi_i\sum\nolimits_{i=0}^\AI i^2\pi_i - (\sum\nolimits_{i=0}^\AI i \pi_i)^2.
\end{aligned}
\right.
\end{equation*}
By Cauchy's Inequality, we have $ Z_{\AI}>0$. If we define
\begin{equation}
\gamma_i = X_{\AI}/Z_{\AI} - i \times Y_{\AI}/Z_{\AI},
\label{eq: gamma_i}
\end{equation}
we have $\gamma_i > 0, \textrm{ for } 0 \leq i \leq \AI$, and
$\gamma_i \leq 0, \textrm{ for }  \AI + 1 \leq i \leq k .$

We can  verify that $(\boldsymbol{\tilde{w}},
\boldsymbol{\tilde{u}}, \tilde{v_1}, \tilde{v_2})$ which is defined
as follows satisfies the KKT conditions (\ref{eq: KKT}). Thus,
$\boldsymbol{\tilde{w}}$ is the global minimum.
\begin{equation*}
\left\{\begin{aligned}
  &\tilde{v}_1 = -2 X_{\AI}/Z_{\AI},\\
&\tilde{v}_2=2Y_{\AI}/Z_{\AI},
\end{aligned}
\right.
\,\,
\left\{\begin{aligned}
      &\tilde{w}_i=\gamma_i, \,\,\,\tilde{u}_i = 0, \,\, 0\leq i \leq \AI,\\
       &\tilde{w}_i=0, \,\,\,\tilde{u}_i = -2 \gamma_i \pi_i, \,\, \AI+1\leq i \leq k,
\end{aligned} \right.
\end{equation*}


Similarly, we can find the optimal solution for the case when
$\AC^*> \sum_{i=0}^{k}i\pi_i$. Since the framework of the proof is
very similar, we only present the solution. In particular, we define
\begin{equation*}
\left\{\!\!\!\begin{aligned}
  X_{\AL}&\!\!=\!\!\sum\nolimits_{i=\AL}^k\! i^2\pi_i \!-\! \mathcal{C^*}\!\!\sum\nolimits_{i=\AL}^k\! i \pi_i,\,\,
  Y_{\AL}\!\!=\!\!\sum\nolimits_{i=\AL}^k\! i\pi_i \!-\! \mathcal{C^*}\!\!\sum\nolimits_{i=\AL}^k \!\pi_i,\\
  Z_{\AL}&=\sum\nolimits_{i=\AL}^k \pi_i\sum\nolimits_{i=\AL}^k i^2\pi_i - (\sum\nolimits_{i=\AL}^k i \pi_i)^2,
\end{aligned}
\right.
\end{equation*}
where $\AL$ is an integer which satisfies the following condition.
\begin{equation}
\AL\!> \!0 \mbox{ and } \AL\! > \!X_{\AL}/Y_{\AL} \!\geq\!
\AL-1, \mbox{ or }
  \AL \!= \!0 \mbox{ and } \AL \!>\! X_{\AL}/Y_{\AL}.
\label{eq: definitionL}
\end{equation}
If we define
\begin{equation}
\Gamma_i  = X_{\AL}/Z_{\AL}- i \times Y_{\AL}/Z_{\AL},
\label{eq: Gammai}
\end{equation}
we can also verify that $(\boldsymbol{\tilde{w}},
\boldsymbol{\tilde{u}}, \tilde{v_1}, \tilde{v_2})$ which is defined
as follows satisfies the KKT conditions. Therefore,
$\boldsymbol{\tilde{w}}$ is the global minimum.
\begin{equation*}
\left\{\begin{aligned}
&\tilde{v}_1 = -2 X_{\AL}/Z_{\AL},\\
&\tilde{v}_2 = 2 Y_{\AL}/Z_{\AL},
\end{aligned}
\right.
\,\,
\left\{\begin{aligned}
&\tilde{w}_i=0,\,\,\,\tilde{u}_i=-2 \Gamma_i \pi_i, \,\, 0 \leq i \leq \AL \!-\!1,\\
&\tilde{w}_i=\Gamma_i,\,\,\,\tilde{u}_i=0, \,\, \AL \leq i \leq k.
\end{aligned}
\right.
\end{equation*}

The cases when $\AC^*=0$ or $k$ and $\AC^*=\sum_{i=0}^{k}i\pi_i$
correspond to the greedy and random algorithms, respectively.
Therefore, the maximum wear-leveling  $\mathcal{W^*}$ can be derived
as in Equation (\ref{eq: maximum wear-leveling})
where $\gamma_i$, $\AI$, $\Gamma_i$, and $\AL$ are defined by
Equations~(\ref{eq: step3})-(\ref{eq: Gammai}). \done

\end{document}